\documentclass[11pt,draftcls,onecolumn]{IEEEtran}
\usepackage{amsmath,amsfonts,amssymb,psfig,color}
\usepackage[breaklinks=true, colorlinks=true, linkcolor=black, urlcolor=dblue,
  citecolor=black, pdfpagemode=None, pdfstartview=FitH]{hyperref}
\definecolor{gray}{cmyk}{.2,0.2,.3,.1}
\definecolor{dred}{cmyk}{0,0.9,0.4,0.3}
\definecolor{dblue}{rgb}{0,0,0.5}
\definecolor{dgreen}{rgb}{0,0.3,0}
\definecolor{dgray}{rgb}{0.3,0.3,0}

\DeclareOldFontCommand{\rm}{\normalfont\rmfamily}{\mathrm}
\DeclareOldFontCommand{\sf}{\normalfont\sffamily}{\mathsf}
\DeclareOldFontCommand{\tt}{\normalfont\ttfamily}{\mathtt}
\DeclareOldFontCommand{\bf}{\normalfont\bfseries}{\mathbf}
\DeclareOldFontCommand{\it}{\normalfont\itshape}{\mathit}
\DeclareOldFontCommand{\sl}{\normalfont\slshape}{\@nomath\sl}
\DeclareOldFontCommand{\sc}{\normalfont\scshape}{\@nomath\sc}

\title{\huge Lattice Quantization with Side Information: Codes,
  Asymptotics, and Applications in Sensor Networks \thanks{S.\ D.\
  Servetto is with the School of Electrical and Computer Engineering,
  Cornell University.  URL: \href{http://cn.ece.cornell.edu/}
  {{\tt http://cn.ece. cornell.edu/}}.  Work supported by the National
  Science Foundation, under awards CCR-0227676, CCR-0238271 (CAREER),
  CCR-0330059, and ANR-0325556.  This paper is based in part on work
  presented at the IEEE Data Compression Conference in
  2000~\cite{Servetto:02b}, and at the Allerton conference in
  2002~\cite{Servetto:02c}.}}
\author{Sergio D. Servetto}
\date{August 31, 2006.}

\begin{document}
\maketitle
\thispagestyle{empty}

\begin{picture}(0,0)
\put(-8,75){\tt To appear in the IEEE Transactions on Information Theory.}
\end{picture}

\vspace{-15mm}
\begin{abstract}
\noindent\it
We consider the problem of rate/distortion with side information
available only at the decoder.  For the case of jointly-Gaussian
source $X$ and side information $Y$, and mean-squared error distortion,
Wyner proved in 1976 that the rate/distortion function for this problem
is identical to the conditional rate/distortion function $R_{X|Y}$,
assuming the side information $Y$ is available at the encoder.  In
this paper we construct a structured class of asymptotically optimal
quantizers for this problem: under the assumption of high correlation
between source $X$ and side information $Y$, we show there exist
quantizers within our class whose performance comes arbitrarily close
to Wyner's bound.  As an application illustrating the relevance of
the high-correlation asymptotics, we also explore the use of these
quantizers in the context of a problem of data compression for sensor
networks, in a setup involving a large number of devices collecting
highly correlated measurements within a confined area.  An important
feature of our formulation is that, although the per-node throughput
of the network tends to zero as network size increases, so does the
amount of information generated by each transmitter.  This is a
situation likely to be encountered often in practice, which allows
us to cast under new---and more ``optimistic''---light some negative
results on the transport capacity of large-scale wireless networks.
\rm
\end{abstract}

\vspace{15mm}
\vspace{15mm}

\setcounter{page}{0}
\pagebreak

\section{Introduction}

\subsection{Large-Scale Wireless Sensor Networks}

Wireless networks span a wide spectrum in terms of their functionality
(i.e., what they are used for), organization (i.e., how the different
components are assembled to form a complete working system), and the
technologies used to build them.  A long-term project currently under
way at Cornell deals with the design and prototyping of networks with
the following defining characteristics:
\begin{itemize}
\item The nodes operate under severe power constraints, support
  relatively large data transfer rates, and their number and density
  is large.
\item Once nodes are deployed, their mobility is very limited (if there
  is any at all).  Instead, the main source of uncontrolled dynamics in
  the network is the temporary failure of individual nodes: this will
  typically happen either due to exhaustion of the power source (and for
  the duration of the ``refueling'' period), or due to variations in the
  wireless medium.
\end{itemize}
In our setup of interest, the network is made up of devices whose
functionality is essentially that of a traditional Cisco router, with
the addition that they communicate over a wireless channel, their size
is many orders of magnitude smaller, and they may come equipped with
sensors that generate information locally as well.  Such networks
would prove extremely useful in a variety of very relevant scenarios,
such as disaster relief operations, military and surveillance applications,
cell-size reduction in cellular networks, environmental monitoring, etc.

The development of a working network of this kind requires solutions
to a number of technical challenges (e.g., routing, flow control,
source and channel coding, power control, modem design, hardware, etc.).
Among all these, of particular interest in this paper is the problem of
source coding, in a scenario in which the data collected by a large number
of sensors is highly correlated.  When network nodes are coupled with
devices that sense a spatial process at different locations (e.g.,
concentration of ozone in the atmosphere, spread of a pathogen/pollutant
agent, temperature of a material, etc.), the measurements collected by
each node will not be independent in general, but instead will be
correlated, with a correlation structure determined by the corresponding
fluid dynamics equations.  Furthermore, the higher the density of nodes
in the network, the higher the correlation in the measurements will be.
Therefore, appropriate source coding capable of removing these dependencies
has the potential to significantly reduce the number of bits to be
transmitted (and therefore the consumption of scarce power resources),
when compared to a coding strategy that treats all measurements as being
independently generated.

\begin{figure}[ht]
\centerline{\psfig{file=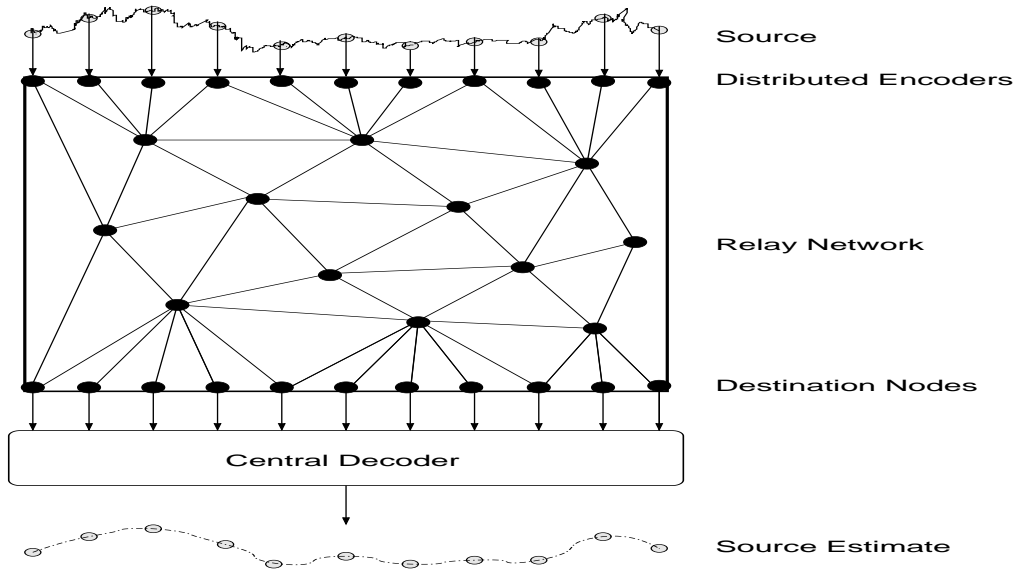,height=7.5cm,width=13.5cm}}
\caption{Network model.  There are three types of nodes: sources,
  relays, and destination nodes, with $n$ nodes of each type.  There is
  a source (a random process whose statistics are known by all sources),
  from which each of the source nodes collects a sample.  These samples are
  encoded by each source node without knowledge of the samples collected by
  other nodes, fed into the network, and each sent to a destination node.
  Finally, these destination nodes pool all their information at a central
  location, at which a decoder forms an estimate of the entire sample path,
  based on the data available from all sensors.  A key aspect of our problem
  formulation is that each source node has to decide what information to send
  to the central decoder {\em without} explicit knowledge of the information
  available at other nodes---only with knowledge of the statistics of that
  correlated data.}
\label{fig:network-model}
\end{figure}

The use of standard and well understood source coding techniques is not
appropriate in the context of highly correlated sources: the use of
classical source codes to remove redundancy in the measurements collected
by different sensors requires that data be pooled at a common node prior
to transmission.  But this pooling action consumes valuable communication
resources itself, thus defeating the very same goal it tries to achieve
(communication efficiency).  Therefore, {\em distributed} source coding
techniques are required, i.e., codes capable of removing correlation
among measurements even in the presence of uncertainty about the exact
value measured at remote locations.
To this end, we define a simple abstraction that captures the essential
properties of this problem.  First, we consider the source of information
to be a random process $(X_s)_{s\in[0,1]}$, defined over a bounded set,
and with {\em continuous} sample paths---continuity is one simple way of
capturing into our model the notion of correlation among measurements
increasing with the number of nodes in a confined area.  This process is
observed by a finite number of sensors, and these observations are to be
communicated over a wireless network, as illustrated in
Fig.~\ref{fig:network-model}.

An important aspect of this problem setup is the fact that, as we
increase the number of source nodes, the amount of information contained
in each sample tends to zero---because the source is continous, two nearby
samples are almost the same.  And we know from recent work on the transport
capacity of one class of wireless networks that, again for large networks,
the per-node throughput of networks in this class also tends to
zero~\cite{GuptaK:00}.  Therefore, {\em provided that the rate at which
information contained in each sample decays at least as fast as the
throughput of the network}, appropriate source coding techniques should
enable an accurate reconstruction of the source at the central decoder
of Fig.~\ref{fig:network-model}.  A study of the resulting source coding
problem in the context of these networks is the central subject of this
paper.

\subsection{Rate Distortion with Side Information}

\subsubsection{Problem Statement}

Let $\{(X_n,Y_n)\}_{n=1}^\infty$ be a sequence of independent drawings
of a pair of dependent random variables $X$ and $Y$, and let $D(x,\hat{x})$
denote a single-letter distortion measure.  The problem of rate distortion
with side information at the decoder asks the question of how many bits
are required to encode the sequence $\{X_n\}$ under the constraint that
${\tt E}D(x,\hat{x}) \leq d$, assuming the side information $\{Y_n\}$ is
available to the decoder but not to the
encoder~\cite[Ch.\ 14.9]{CoverT:91}.  This problem, first
considered by Wyner and Ziv in~\cite{WynerZ:76}, is a special
case of the general problem of coding correlated information sources
considered by Slepian and Wolf~\cite{SlepianW:73b},
in that one of the sources ($\{Y_n\}$) is available {\em uncoded} at the
decoder.  But it also generalizes the setup
of~\cite{SlepianW:73b}, in that coding is with respect
to a fidelity criterion rather than noiseless.  One important motivation
for us to consider this problem is the fact that good quantizers with side
information will be used in the proof of scalability of a large sensor
network.

In~\cite{Wyner:78, WynerZ:76}, Wyner and
Ziv derive the rate/distortion function $R^*(d)$ for this problem, for
general sources and general (single letter) distortion metrics.  In this
work however we restrict our attention only to Gaussian sources, and mean
squared error (MSE) distortion.  This case is of special interest because,
under these conditions, it happens that $R^*(d) = R_{X|Y}(d)$, the
conditional rate/distortion function {\em assuming $Y$ is available at the
encoder}~\cite{Wyner:78, WynerZ:76}.  We
are intrigued by the fact that there exist coding methods which can perform
as well as if they had access to the side information at the encoder, even
though they don't.  One goal pursued in this paper then is the construction
a family of quantizers which realizes these promised gains.

\subsubsection{Lattice Quantization with Side Information}

High-rate quantization theory provides much of the motivation to consider
lattices~\cite{GrayN:98}. Under an assumption of fine
quantization, the performance of an $n$-dimensional quantizer $\Lambda$
whose Voronoi cells are all congruent to a polytope $P$ is given by
\begin{equation}
   d = G(P) \cdot e^{-\frac{2}{n}({\cal H}(\Lambda,p_X)-h(p_X))},
   \label{eq:zador-gersho-bound}
\end{equation}
where $p_X$ is the joint source distribution in $n$ dimensions, ${\cal H}$
is the discrete entropy induced on the codebook $\Lambda$ by quantization of
the source $p_X$, $h$ is the differential entropy, and
\[ G(P) = \frac{\frac{1}{n}
            \int_P ||{\bf x}-{\bf \hat{x}}||^2 \mbox{ d\bf x}}
            { \left(\int_P \mbox{ d\bf x}\right)^{1+\frac{2}{n}} }
\]
is the normalized second moment of $P$ (using MSE as a distortion
measure)~\cite{gersho:quantization-asymptotics,zador:quantization-asymptotics}.

In the problem of rate distortion with side information, for Gaussian
sources and MSE distortion, the goal is to attain a distortion value
$d$ using $R_{X|Y}(d) < R_X(d)$ nats/sample.  In~(\ref{eq:zador-gersho-bound})
this means that, at fixed bit rate $R_0$, we want to design quantizers
that achieve distortion
\[ d_0 \approx c_n \cdot e^{-\frac{2}{n}(nR_0-h(p_{X|Y}))} \]
when coding $X$, where $c_n \leq G(P)$ is the coefficient of quantization
in $n$ dimensions~\cite{gersho:quantization-asymptotics}.  But since we do not
have access to $Y$ (we only know $p_{X|Y}$), using classical quantizers we can
only attain a distortion value
\[ d \approx c_n \cdot e^{-\frac{2}{n}(nR_0-h(p_X))} > d_0 \]
(because {\small $h(X|Y) < h(X)$}), or equivalently, we need to use
some extra rate $\rho \approx R_X-R_{X|Y}$ such that
\[ d_0 \approx c_n \cdot  e^{-\frac{2}{n}(n(R_0+\rho)-h(p_X))}. \]
What makes this problem interesting is that we are only allowed to use $R_0$
nats/sample, not $R_0+\rho$.  One way to do that has been proposed by Shamai,
Verd\'{u} and Zamir in~\cite{shamai-verdu-zamir:systematic-lossy-coding,
zamir-shamai:almost-there}, which consists of: (a) taking a codebook with
roughly $e^{n(R_0+\rho)}$ codewords and distortion $d_0$, (b) partitioning
this codebook into $e^{nR_0}$ sets of size $e^{n\rho}$ each, (c) encoding
only enough information to identify each one of the $e^{nR_0}$ sets, and
(d) using the side information $Y$ to discriminate among the $e^{n\rho}$
codewords collapsed into each set.  One of our motivations for considering
lattice codes is the fact that their structure makes it particularly easy
to express these partitioning operations described
in~\cite{shamai-verdu-zamir:systematic-lossy-coding}.

We should also mention that another reason to consider lattices is our
wish to answer a challenge posed by Zamir and Shamai
in~\cite{zamir-shamai:almost-there}.  They present an encoding procedure
very closely related to the one we propose here, they argue the existence
of good lattices to use with that procedure, they study their distortion
performance, but they do not present any examples of concrete
constructions: their paper concludes by saying that (sic) ``{\em beyond
the question of existence, it would be nice to find specific constructions
of good nested codes}''.  Finding those specific constructions is one of
the original contributions in this work.

\subsection{Related Work}

Note: this section contains relevant related work as of Fall 2004.

\subsubsection{Codes and Quantizers}

The design of quantizers for the problem of rate distortion with side
information was considered recently by Shamai, Verd\'{u} and Zamir, where
they present design criteria for two different cases: Bernoulli sources
with Hamming metric, and jointly Gaussian sources with mean squared error
metric~\cite{shamai-verdu-zamir:systematic-lossy-coding,
zamir-shamai:almost-there}.  The key contribution presented in that work
is a constructive mechanism for, given a codebook, using the side
information at the decoder to reduce the amount of information that needs
to be encoded to identify codewords, while at the same time achieving
essentially the distortion of the given codebook.  That work provided
much inspiration for our work on the design of lattice codes presented in
this paper.

Other work on code constructions includes the application of similar
codebook partitioning ideas in the context of trellis
codes~\cite{sandeep-kannan:discus}, a preliminary version of this
work~\cite{Servetto:02b}, generalizations to the case when the side
information may be coded as well~\cite{PradhanR:00,ZhaoE:01},
constructions based on LDPC codes~\cite{AaronG:02, MitranB:02,
TianGZ:03}, and other code constructions~\cite{LiuCLX:04,
RebolloMonederoZG:03}.

\subsubsection{Information-Theoretic Performance Bounds}

Whereas there has been some interest in recent times on the more
practical aspects of these problems, a significant amount of work on
related topics had already been done before in the context of multiuser
information theory.  Specifically on the problem of rate/distortion
with side information, besides the above mentioned work of Wyner and
Ziv~\cite{Wyner:78, WynerZ:76}, Kaspi
and Berger present a summary of known results and a number of new
results (as of 1982) in~\cite{KaspiB:82}, leaving only a couple of
special cases still open.  Heegard and Berger further generalize to the
case when there is uncertainty on whether the side information is available
at the decoder or not~\cite{heegard-berger:uncertain-side-info}.  For
an arbitrary pair of sources, Zamir gives bounds on how far away the
conditional rate/distortion function and the Wyner-Ziv rate/distortion
function can be from each other~\cite{Zamir:96}.

Closely related to the problem of rate/distortion with side information
is that of {\it Noiseless Coding of Distributed Correlated Sources}.
Slepian and Wolf formulate this problem, and
determine the minimum number of bits per symbol required to encode two
correlated sequences $\{X_n\}$ and $\{Y_n\}$ separately, such that they
can be faithfully reproduced by a centralized decoder, under the assumption
that $\{(X_n,Y_n)\}_{n=1}^\infty$ is
i.i.d.~\cite{SlepianW:73b}.  Cover then gives a simpler
proof of the same result, which also generalizes to arbitrary ergodic
processes, countably infinite alphabets, and arbitrary number of correlated
sources~\cite{Cover:75b}.  Wyner presents an information theoretic
characterization of the minimum rates required for faithful reproduction in
a general network with side information~\cite{Wyner:75}.  Barros and
Servetto consider the Slepian-Wolf problem in an arbitrary network
setup with noisy point-to-point links~\cite{BarrosS:06}.

A long-standing open problem in network information theory is the
characterization of the rate-distortion region for the {\em Multiterminal
Source Coding} problem, which is basically the Slepian-Wolf problem,
but in which a non-zero distortion is allowed in the encoding of
both sources.  The most significant contribution to this date can be
found in Tung's doctoral dissertation~\cite{Tung:PhD}.  Berger
developed some useful notes for
a tutorial lecture on this and related problems~\cite{Berger:78}.

Yet another closely related problem is {\it the CEO Problem}.  In this
version, multiple sensors observe
{\em noisy} versions of the same signal, and must convey their observations
to a centralized decoder at a combined rate of not more than $R$ bits/sample.
This case generalizes the problem of encoding correlated observations,
to the case when the number of sensors is large, and to the case when the
signal to be communicated cannot be observed directly.  Berger et al.\
present a solution to this problem in the general case~\cite{BergerZV:96}.
Viswanathan and Berger specialize the results of~\cite{BergerZV:96} to the
Quadratic-Gaussian case~\cite{ViswanathanB:97}: an interesting conclusion
in this case is that the optimal rate of decay of the error is of the form
$R^{-1}$ when the sensors cannot communicate prior to transmission, as
opposed to an exponential decay otherwise.

An interesting duality between the problem of rate/distortion with side
information discussed above, and the problem of channel coding with side
information at the transmitter~\cite{Costa:83}, has been pointed out by
several groups~\cite{BarronCW:02,PradhanCR:03,SuEG:00}.  Cover and Chiang
present a comprehensive coverage of duality issues in problems with side
information~\cite{CoverC:02}, and Chiang and Boyd fully develop an
optimization-theoretic approach to analyzing the duality of channel
capacity and rate distortion problems~\cite{ChiangB:04}.  Merhav and
Shamai established a separation theorem in this context~\cite{MerhavS:03}.
Therefore, it should be possible to derive good codes for one problem
from good codes available for the other.

Zamir et al.\ present a very interesting tutorial on noisy multiterminal
networks, with many useful references~\cite{ZamirSE:02}.

\subsubsection{Performance of Wireless Networks}

A key result in the analysis of performance of wireless networks states
that when $n$ non-mobile nodes are optimally placed in a disk of unit area,
traffic patterns are optimally assigned, and the range of each transmission
is optimally chosen, the total throughput that the network can carry is
$O(\sqrt{n})$~\cite{GuptaK:00}.  As a result, the per-node throughput is
only $O(\frac{1}{\sqrt{n}})$, i.e., decays to zero as the number of nodes
in the network increases.  Other results along the same lines were presented
in~\cite{GuptaK:03, XieK:04}.

The work of~\cite{GuptaK:00} sparked significant interest in this problem.
When nodes are allowed to move, assuming transmission delays proportional
to the mixing time of the network, the total network throughput is $O(n)$,
and therefore the network can carry a non-vanishing rate per
node~\cite{GrossglauserT:02}.  Using a linear programming formulation,
non-asymptotic versions of the results in~\cite{GuptaK:00} are given
in~\cite{ToumpisG:02}.  Using pure network flow methods, similar results
(and generalizations thereof) have been obtained
in~\cite{PerakiS:03, PerakiS:04}.  An alternative method for deriving
transport capacity was presented in~\cite{KulkarniV:04}.

\subsection{Main Contributions and Organization of the Paper}

This paper presents the following original contributions:
\begin{itemize}
\item The construction of lattice codes for the problem of rate/distortion
  with side information.  We propose a design procedure based on the choice
  of a lattice that is a good quantizer for the classical rate/distortion
  problem, and a geometrically-similar sublattice, inspired by the idea of
  partitioning codebooks to obtain good codes for this problem proposed
  in~\cite{shamai-verdu-zamir:systematic-lossy-coding,
  zamir-shamai:almost-there}, and by our previous work on the design of
  lattice quantizers for multiple description coding~\cite{VaishampayanSS:01}.
\item An asymptotic analysis (in rate and correlation) of the performance
  of these codes which, to the best of our knowledge, is the first such
  analysis for Wyner-Ziv codes.  Our analysis reveals some interesting
  shortcomings of these codes, and suggest a simple modification to make
  to the construction to ensure their optimality.  These optimal codes
  effectively answer a challenge of Zamir and
  Shamai~\cite{zamir-shamai:almost-there}.
\item The illustration that high correlation asymptotics in source coding
  are indeed a new asymptotic regime with very meaningful practical
  implications.  So far source coding has considered two asymptotic
  regimes: large block asymptotics~\cite{Shannon:59}, or high
  rate asymptotics~\cite{zador:quantization-asymptotics}.  High correlation
  asymptotics are a new asymptotic regime that, as we will see in
  Section~\ref{sec:sensor-networks}, proves quite relevant in the context
  of new problems derived from sensor networking applications.
\item The identification of a large class of applications for which the
  vanishing rates property of wireless networks does not pose a problem,
  by virtue of the fact that the amount of information that each node needs
  to transmit decays at the same rate as (or faster than) throughput does.
\end{itemize}

The rest of this paper is organized as follows.  In
Section~\ref{sec:code-design} we present the structure of lattice
quantizers for the problem of rate/distortion with side information,
and in Section~\ref{sec:asymptotics} we evaluate the performance of
the codes obtained, under the assumption of high-correlation between
the source $X$ and the side information $Y$.  In
Section~\ref{sec:sensor-networks} we illustrate how the proposed
codes can be used to deal effectively with the vanishing rates
property of an important class of large-scale sensor networks.
Final remarks are presented in Section~\ref{sec:conclusions}.

\section{Design of Lattice Codes with Side Information}
\label{sec:code-design}

\subsection{Definitions}

A source generates a sequence of zero-mean iid pairs
$(x_i,y_i)_{i=0}^\infty$, with jointly Gaussian distribution
\[
  f_{X,Y}(x,y) = \frac{1}{2\pi\sigma_X\sigma_Y\sqrt{1-\rho^2}}\;\;
                 e^{-\frac{1}{2(1-\rho^2)}
                    \left(\frac{x^2}{\sigma_X^2}
                          -\frac{2\rho x y}{\sigma_X\sigma_Y}
                          +\frac{y^2}{\sigma_Y^2}\right)},
\]
with covariance matrix ${\bf K} = {\tiny \left[\!\begin{array}{cc}
\sigma_X^2 & \rho\sigma_X\sigma_Y \\
\rho\sigma_X\sigma_Y & \sigma_Y^2 \\
\end{array}\!\right]}$, and correlation coefficient $\rho$.  The corresponding
conditional and marginal densities are denoted by $f_{Y|X}$, $f_{X|Y}$,
$f_X$, $f_Y$.  For a set of $n$ linearly independent column vectors
$\{{\bf v}_1,...,{\bf v}_n\}$, a {\em lattice} $\Lambda\subset\mathbb{R}^n$
is defined by
\[
   \Lambda = \left\{ \sum_{i=1}^n c_i {\bf v}_i : c_1...c_n\in\mathbb{Z}
             \right\},
\]
and its {\em generator matrix} ${\bf V}=\left[{\bf v}_1|...|{\bf v_n}\right]$.
The volume of a polytope $P\subset\mathbb{R}^n$ is denoted by $\nu(P)$.
For a constant $s\in\mathbb{R}$, the {\em scaled lattice} $s\Lambda$ is the
lattice generated by $s{\bf V}$, where ${\bf V}$ is the generator matrix of
a lattice $\Lambda$.  The {\em Voronoi cell} of a lattice point $\lambda$ in
the lattice $\Lambda$ is defined by
\[
   V[\lambda\!:\!\Lambda]
     = \{{\bf x}\in\mathbb{R}^n:||{\bf x}-\lambda||^2\leq||{\bf x}-\lambda'||^2,
         \;\forall\lambda'\in\Lambda \}.
\]
The {\em nearest neighbor map of a lattice} is a function
$Q_\Lambda : \mathbb{R}^n \rightarrow \Lambda$, defined by
\[
   Q_\Lambda({\bf x}) = \arg\min_{\lambda\in\Lambda} ||{\bf x}-\lambda||^2,
\]
where ties are broken arbitrarily (e.g., numbering all the $\lambda$'s,
and assigning ${\bf x}$ to the $\lambda$ with smallest index).  From the
definitions it follows trivially that $V[\lambda\!:\!\Lambda] =
\{{\bf x}\in\mathbb{R}^n:Q_\Lambda({\bf x})=\lambda\}$, except possibly
for a set of measure zero.  A lattice $\Lambda'$ is a {\em sublattice} of
a lattice $\Lambda$ if $\Lambda'\subseteq\Lambda$.  The {\em quotient
group}~\cite[Sec.\ 6.3]{Bourbaki:58} of a lattice modulo a sublattice is
denoted by $\Lambda/\Lambda'$, and its order by $|\Lambda/\Lambda'|$.

A {\em Wyner-Ziv Lattice Vector Quantizer} (WZ-LVQ) is a triplet
${\cal Q}=(\Lambda,\kappa,s)$, where:
\begin{itemize}
\item $\Lambda$ is a lattice.
\item $\kappa: \mathbb{R}^n \rightarrow \mathbb{R}^n$ is a linear operator
      such that $\kappa{\bf u}\cdot\kappa{\bf v} = c\;{\bf u}\cdot{\bf v}$
      (for some $c>0$), and such that $\kappa(\Lambda) \subseteq \Lambda$.
      Essentially, $\kappa$ defines a {\em similar} sublattice of
      $\Lambda$.\footnote{Two lattices $\Lambda_1$, $\Lambda_2$ (with
      generator matrices $M_1$, $M_2$) are said to be {\em similar} when
      there is a constant $c \neq 0$, an integer matrix U with
      $|\mbox{det}(U)| = 1$, and a real matrix $B$ with $BB^{\top} = I$,
      such that $M_2 = c \; U M_1 B$~\cite{neil:splag}.
      Intuitively, similar lattices ``look the same'', up to a rotation,
      a reflection, and a change of scale.}
\item $s \in (0,\infty)$ is a scale factor that expands (or shrinks)
      $\Lambda$ and $\kappa(\Lambda)$.
\end{itemize}

Intuitively, the lattice $\Lambda$ is the fine codebook, the one whose
codewords are to be partitioned into equivalence classes.  We choose to
implement this partition by considering a sublattice $\Lambda' \subseteq
\Lambda$, and then considering the resulting quotient group $\Lambda/\Lambda'$.
$s$ is a constant that multiplies the generator matrices of the lattices
considered, which is to be adjusted as a function of the correlation
between the source $X$ and the side information $Y$.  A justification for
the choice of a {\em similar} sublattice (as opposed to any other sublattice)
to implement the codebook partition, and a justification for the explicit
introduction of a scale factor $s$ as a parameter of the quantizer (as
opposed to having this lattice scale be determined by the coding rate, as
in classical quantization theory) will become apparent later, after we study
the rate-distortion performance of the proposed quantizers.

The question of the existence of similar sublattices arose in connection
with another vector quantization problem~\cite{VaishampayanSS:01}, and also
in the study of symmetries of
quasicrystals~\cite{baake-moody:similarity-submodules-semigroups}.  The
subject is thoroughly covered in~\cite{conway-rains-neil:similar-sublattices},
where necessary (and in some cases sufficient) conditions are given for
their existence.

\subsection{Encoding/Decoding Algorithms}

\begin{figure}[ht]
\centerline{\psfig{file=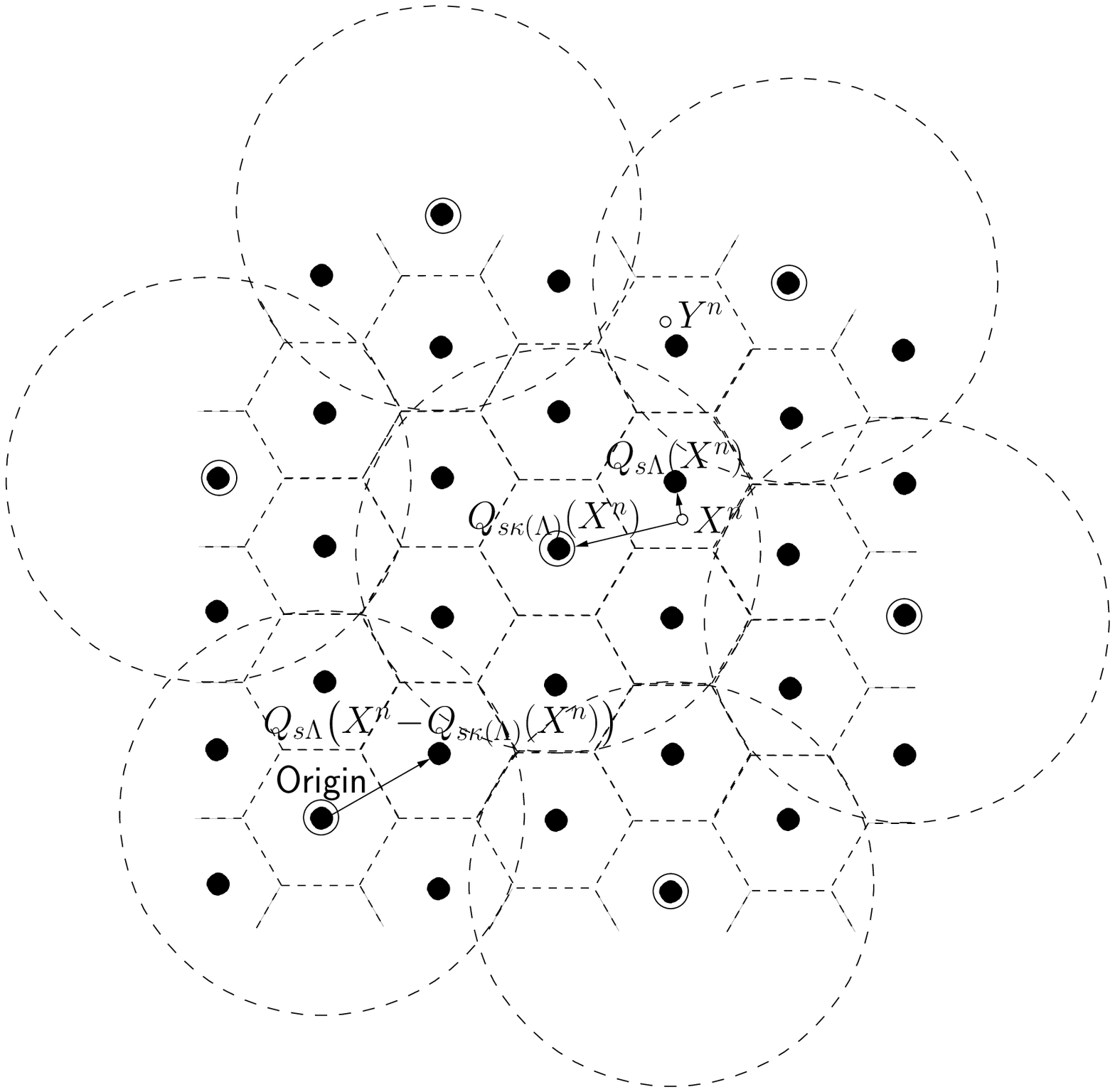,height=10cm}
            \psfig{file=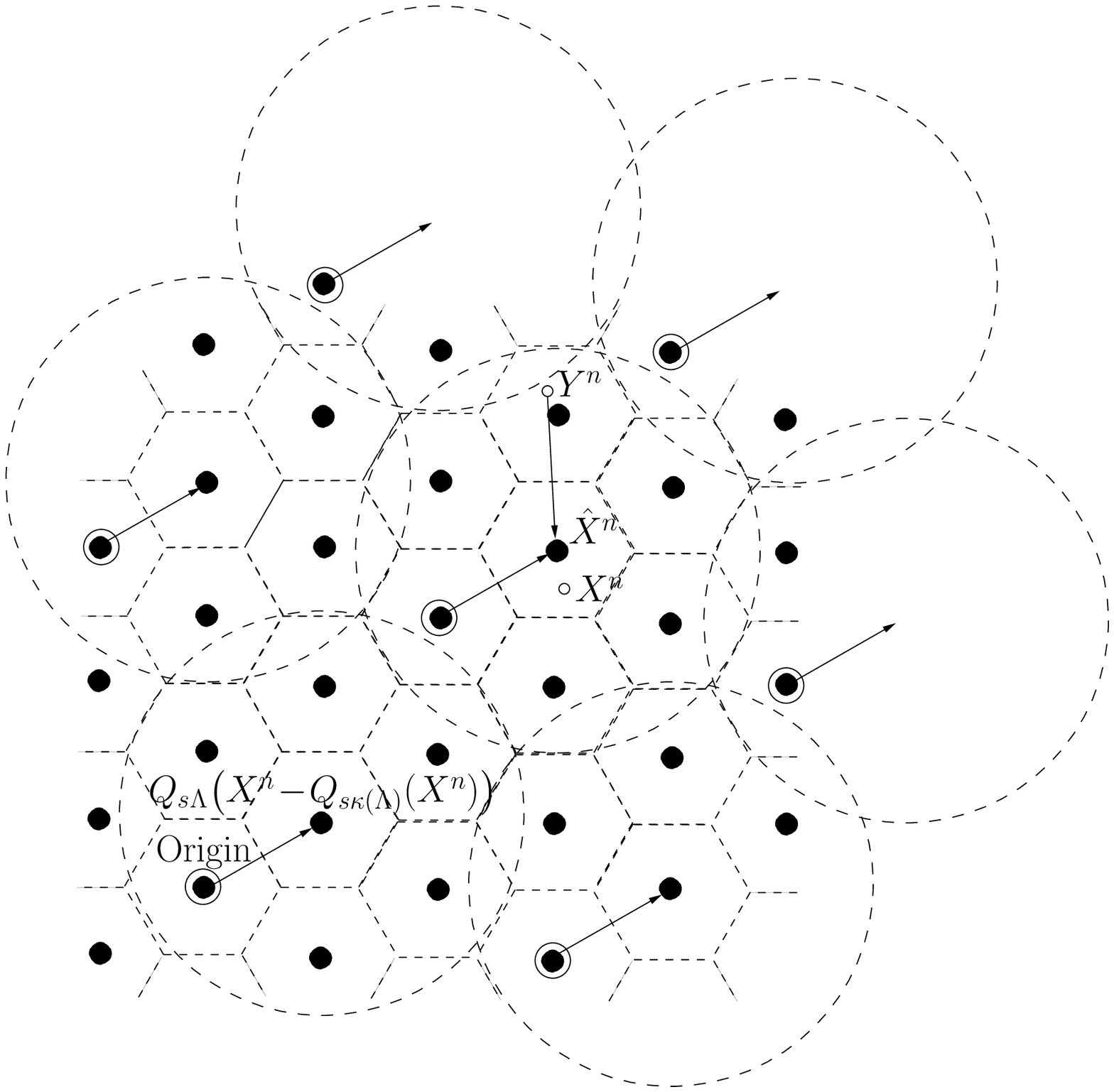,height=10cm}}
\vspace{-2mm}
\caption{To illustrate the mechanics of the proposed quantizers
  (left: encoding, right: decoding).  A sublattice similiar to the base
  lattice is chosen (circled points), matched to how far $X^n$ and $Y^n$
  are expected to be: in this example, with high probability $X^n$ and
  $Y^n$ are in neighboring Voronoi cells of the fine lattice.  Then
  $X^n$ is quantized first with the coarse lattice, then this coarse
  description is subtracted from $X^n$, and this difference is quantized
  again with the fine lattice; this quantized difference is then sent to
  to the decoder, as a representative of the set of all codewords
  collapsed into the same equivalence class.
  At the decoder, the entire class is recreated (all the points with a
  thick arrow in the right picture), and among these, the point closest
  to the side information $Y^n$ is declared to be the original quantized
  value for $X^n$.  Note that there is always a chance that a particular
  realization of the noise process may take $Y^n$ too far away from
  $X^n$, in which case a decoding error occurs.}
\label{fig:encoder-decoder}
\end{figure}

Let $X^n$ denote a block of $n$ source samples, and $Y^n$ a block of $n$
side information samples.  The encoder and decoder are maps
$f_n:\mathbb{R}^n \rightarrow s\Lambda/s\kappa(\Lambda)$ and
$g_n:s\Lambda/s\kappa(\Lambda)\times\mathbb{R}^n \rightarrow s\Lambda$,
defined by
\begin{equation}
   f_n(X^n) = Q_{s\Lambda}\big(X^n-Q_{s\kappa(\Lambda)}(X^n)\big),
   \hspace{1cm}
   \hat{X}^n = g_n(f_n(X^n),Y^n) = Q_{s\kappa(\Lambda)+f_n(X^n)}(Y^n),
  \label{eq:q-alg}
\end{equation}
whose operation is illustrated in Fig.~\ref{fig:encoder-decoder}, with an
example based on the lattice $A_2$.

\subsection{Rate Computation}
\label{sec:rate-computation}

There are only $N = |\Lambda/\kappa(\Lambda)|$ possible different quantizer
outputs, each one with probability $p_k$ ($k=1...N$) given by
\[
   p_k \;=\; \sum_{\lambda\in s\Lambda}
         \int_{V[\kappa(\lambda)+\gamma_k:s\Lambda]}
         f_X({\bf x}) \mbox{ d\bf x},
\]
where $\gamma_k \in s\Lambda/s\kappa(\Lambda)$, and where we identify
the entire equivalence class with a canonical representative taken from
$\Lambda \; \cap \; V[{\bf 0}\!:\!\kappa(\Lambda)]$.  The rate of a
quantizer is then given by
\[ R \;=\; \mbox{$\frac 1 n$} \sum_{k=1}^N p_k \ln(1/p_k), \]
expressed in units of nats per source sample.

Assume now, as is standard in fine-resolution quantization theory, that
Voronoi cells of the quantizers under consideration are small.  In this
case, this translates into a requirement for {\em sublattice} cells to
be small, for which we have that
\[
   \nu(s\kappa(\Lambda)) \; = \; s^n \nu(\kappa(\Lambda))
     \; = \; s^n \nu(N^{\frac 1 n}U\Lambda) \; = \; s^nN \nu(\Lambda)
     \; = \; s^nN,
\]
where the second equality follows from the fact that
$N = |\Lambda/\kappa(\Lambda)| = c^{\frac n 2}$, where $c$ is the norm
of the similarity defined by
$\kappa$~\cite{conway-rains-neil:similar-sublattices} (and therefore
the corresponding scaling is $\sqrt{c}$), $U$ is unitary, and the last
equality follows from assuming $\Lambda$ is normalized to have determinant
1~\cite{neil:splag}.  Then, we see that requiring small
sublattice cells translates into requiring that $s^nN$ be a small
number.  Now, under this assumption, the rate expression above admits a
much simpler form:

\[
1 = \sum_{\lambda\in s\Lambda}
           \int_{V[\lambda:s\Lambda]} f_X({\bf x})\mbox{ d\bf x} \\
  = \sum_{\gamma_k\in s\Lambda/s\kappa(\Lambda)}
           \underbrace{\sum_{\lambda\in s\Lambda}
            \int_{V[\kappa(\lambda)+\gamma_k:s\Lambda]}
               f_X({\bf x})\mbox{ d\bf x}}_{p_k}.
\]
The integral of the source density in $p_k$ can be approximated by
\[
  f_X(\kappa(\lambda)+\gamma_k)\;\cdot\;
                \nu(V[\kappa(\lambda)+\gamma_k:s\Lambda]).
\]
But assuming small cells
for the sublattice (standard in quantization theory), since the Gaussian
source is continuous, we have that within a cell of $\kappa(\Lambda)$ $f_X$
is approximately constant, and hence independent of the particular shift
$\gamma_k$.  Furthermore, since $\Lambda$ is a lattice, all its cells are
congruent, and therefore their volumes are all the same, thus making $\nu$
also independent of the particular shift $\gamma_k$.  Call $p$ this
(approximately) constant value for $p_k$.  Therefore, we have
\[
   1 \;\; \approx \sum_{\gamma_k\in\Lambda/\kappa(\Lambda)} p
     \;\; = \;\; |\Lambda/\kappa(\Lambda)| p,
\]
and hence,
\begin{eqnarray*}
  p_k \approx \frac{1}{|\Lambda/\kappa(\Lambda)|}
  & \hspace{1cm} \mbox{and} \hspace{1cm}
  & R \approx \mbox{$\frac 1 n$} \log_2 |\Lambda/\kappa(\Lambda)|,
\end{eqnarray*}
independent of $s$ and $f_X$, where the approximations are tight in
the limit as $s^nN\to 0$.

Note that, unlike in classical quantization theory, here the
rate of a quantizer seems to be independent of the size of its Voronoi
cells.  In our context, a high-rate assumption translates into a large
value for $|\Lambda/\kappa(\Lambda)|$, i.e., cells in the fine lattice
are small {\em relative} to the size of cells in the coarse lattice.
But the parameter $s$, which determines the {\em absolute} the size of
these cells, is not part of the rate expression.

\subsection{Distortion Computation}
\label{sec:distortion-nonasymptotic}

Let $\gamma_k({\bf x})$ denote the encoding of a source sequence
${\bf x}$ ($k=1...N$), and $\gamma({\bf x},{\bf y})$ denote the
reconstruction codeword for a source sequence ${\bf x}$ with side
information ${\bf y}$.  Then:
\begin{eqnarray}
\bar d
  & \stackrel{(a)}{=} &
        \mbox{$\frac 1 n$} \int_{{\bf x}\in\mathbb{R}^n} \int_{{\bf y}\in\mathbb{R}^n}
        ||{\bf x} - \gamma({\bf x},{\bf y})||^2 f_{XY}({\bf x},{\bf y})
        \mbox{d}{\bf x}\mbox{d}{\bf y} \nonumber \\
  & = & \mbox{$\frac 1 n$}
        \int_{{\bf x}\in\mathbb{R}^n}\left[\int_{{\bf y}\in\mathbb{R}^n}
        ||{\bf x} - \gamma({\bf x},{\bf y})||^2 f_{Y|X}({\bf y}|{\bf x})
        \mbox{d}{\bf y}\right] f_X({\bf x})\mbox{d}{\bf x} \nonumber \\
  & \stackrel{(b)}{=} & \mbox{$\frac 1 n$} \int_{{\bf x}\in\mathbb{R}^n}
        \left[\sum_{\lambda\in s\kappa(\Lambda)+\gamma_k({\bf x})}
              \int_{{\bf y}\in V[\lambda:s\kappa(\Lambda)+\gamma_k({\bf x})]}
        ||{\bf x} - \lambda||^2 f_{Y|X}({\bf y}|{\bf x})
        \mbox{d}{\bf y}\right] f_X({\bf x})\mbox{d}{\bf x} \nonumber \\
  & \stackrel{(c)}{=} & \mbox{$\frac 1 n$} \int_{{\bf x}\in\mathbb{R}^n}
        \left[\sum_{\lambda\in s\kappa(\Lambda)+\gamma_k({\bf x})}
        ||{\bf x} - \lambda||^2 {\tt Pr}\big({\bf y}\in
        V[\lambda:s\kappa(\Lambda)+\gamma_k({\bf x})]\big|{\bf x}\big)
        \right] f_X({\bf x})\mbox{d}{\bf x} \nonumber \\
  & \triangleq & \mbox{$\frac 1 n$} \int_{{\bf x}\in\mathbb{R}^n}
        \partial({\bf x}, s\kappa(\Lambda)+\gamma_k({\bf x}))
        f_X({\bf x})\mbox{d}{\bf x},
        \label{eq:distortion}
\end{eqnarray}
where:
\begin{itemize}
\item[\small (a)] is just the definition of average distortion;
\item[\small (b)] follows from, for each possible source sequence ${\bf x}$,
partitioning the set of all side information vectors ${\bf y}$ into
Voronoi cells of the sublattice $s\kappa(\Lambda)$, centered at location
$\gamma_k({\bf x})$;
\item[\small (c)] follows from the fact that $||{\bf x}-\lambda||^2$ can
be taken out of the integral, and what remains is an integral of the
conditional density function.
\end{itemize}
The last
definition is introduced to highlight the concept that in quantization
with side information, an entire sublattice plays the role of a single
codeword in classical quantization -- the average error in reconstructing
${\bf x}$ is seen to take the form of an expectation of a suitably
defined distortion metric between source sequences and sublattices.
In Section~\ref{sec:asymptotics} we study the asymptotic behavior
of~(\ref{eq:distortion}), assuming high correlation between $X^n$
and $Y^n$.

\subsection{On the Choice of Similar Sublattices}

As we will see in Section~\ref{sec:asymptotics}, there are some
drawbacks to implementing quantizers for the Wyner-Ziv problem with
a fine quantizer that is essentially a truncated lattice, as follows
from the construction given here.  But there are also significant
benefits to doing so, in terms of the simplicity of this implementation.
So for the time being, if we are going to use two lattices, it is
of interest to consider what kind of lattices should be used.

Suppose we fix the scale factor $s$, and the code rate $\frac{1}{n}\ln(N)$.
Among all the sublattices of $\Lambda$ of index $N$, are there differences
in terms of their distortion performance?  Which sublattices should we
choose?  It follows from~(\ref{eq:distortion}) that a sensible design
criteria is to choose the sublattice which results in maximizing
${\tt Pr}\left\{{\bf y}\in V[{\bf 0}\!:\!s\kappa(\Lambda)]\mid
X\!={\bf x}\right\}$, for ${\bf x}\in V[{\bf 0}\!:\!s\Lambda]$.

\begin{figure}[ht]
\centerline{\psfig{file=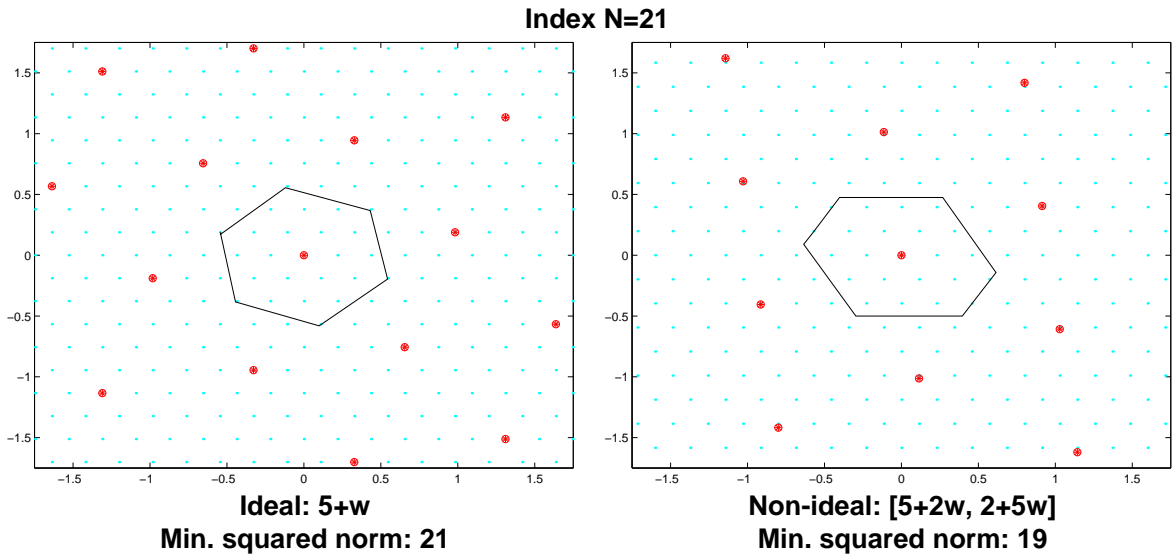,height=7.3cm}}
\vspace{-2mm}
\caption{Two different sublattices of $A_2$, of index $N=21$.  $A_2$
  is isomorphic to the ring of Eisenstein integers
  $\mathbb{Z}(\omega) = \{ a+b\omega\;:\;a,b\in\mathbb{Z};\;
  \omega=[-\frac{1}{2},\frac{\sqrt{3}}{2}]=e^{2\pi i/3}\}$, and {\em ideal}
  sublattices refer to ideals of this ring.  Observe that the ideal sublattice
  of the example has shortest vectors of norm 21, whereas in the non-ideal
  sublattice the shortest vectors are shorter.}
\label{fig:why-similar-sublattices}
\end{figure}

Since the vectors $X$ and $Y$ are jointly Gaussian and with iid
components, the vector $Y|X\!=\!{\bf x}$ is also Gaussian and with iid
components (although the $x_i$'s and the $y_i$'s are certainly not
independent of each other).  The pdf of $Y|X\!=\!{\bf x}$ is therefore
circularly symmetric, and it follows from classical arguments of coding
for Gaussian channels that, to maximize ${\tt Pr}({\bf y}\in V)$, we need
to maximize the norm of the shortest vectors in $\kappa(\Lambda)$.  This
situation is illustrated in Fig.~\ref{fig:why-similar-sublattices}, with
an example based on the lattice $A_2$.

The choice of $A_2$ for illustration purposes in
Fig.~\ref{fig:why-similar-sublattices} is not arbitrary.  In that
particular case, it is known that the minimal norm $\mu$ of any sublattice
of index $N$ in $A_2$ satisfies $\mu \leq N$, and that $\mu = N$ if and
only if the sublattice is ideal~\cite{bernstein-neil-pew:sublattices-of-a2}.
Furthermore, in two dimensions, $A_2$ is both the best classical quantizer
and the best channel coder~\cite{neil:splag}.  Therefore, it seems clear
that a hexagonal lattice and a similar sublattice are the best design
choices in two dimensions: this combination simultaneously minimizes
quantization error, and minimizes the probability of a source vector being
decoded to an incorrect codeword.

Another interesting example is that of very high dimensional spaces.
In this case, we know that good quantizers have (nearly) spherical Voronoi
cells.  But at the same time, spherical cells maximize the minimum distance
between sublattice points, and therefore an optimal sublattice will have
to be similar to the base lattice.

In between dimensions 2 and $\infty$, we are not able to make equally
strong statements---but we use the insights derived from these extreme
cases (a lattice with small second-order moment and a similar sublattice)
as guiding principles, to curb the complexity of the design task.

\section{Asymptotics of Quantizers with Side Information}
\label{sec:asymptotics}

\subsection{Modeling Assumptions and Performance Metric}

\subsubsection{Modeling Assumptions}

\begin{figure}[ht]
\centerline{\psfig{file=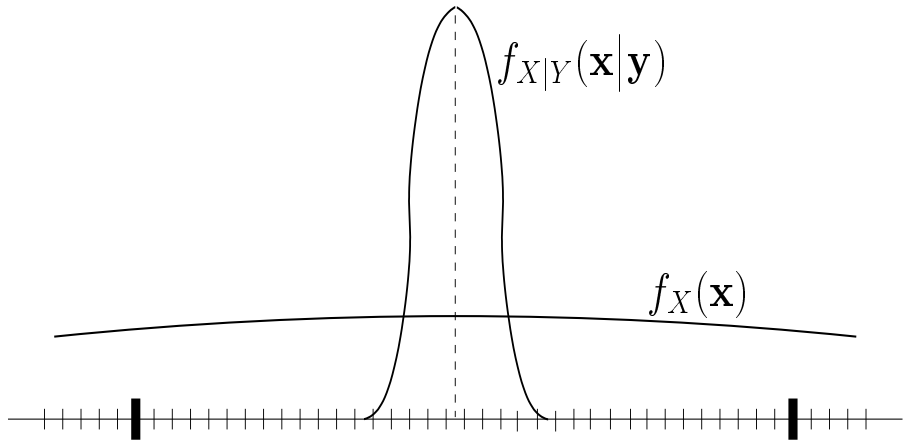,height=6cm,width=12cm}}
\vspace{-2mm}
\caption{Illustration (in one dimension) of the meaning of the asymptotic
  regime considered in this work.  Working under an assumption of high
  correlations, we have that the conditional distribution of the source
  ${\bf x}$ given side information ${\bf y}$ is sharply concentrated around
  its mean value ${\bf y}$ -- as a result, we can make the probability of
  the source ${\bf x}$ away from ${\bf y}$ by more than any positive
  constant be arbitrarily small (by choosing $\rho$ close enough to 1),
  and hence we can assume that sublattice cells, while being vanishingly
  small themselves ($s\approx 0$), can be considered large enough to
  contain most of the probability in $f_{X|Y}$.  Then, because we take
  $R$ large, we further partition each sublattice cell into a large
  number of much smaller fine lattice cells.}
\label{fig:assumptions}
\end{figure}

Our goal in this section is to find a simpler expression for $\bar{d}$
than that presented in Section~\ref{sec:distortion-nonasymptotic}.  To
do so, we work under some extra assumptions:
{\it\begin{itemize}
\item The correlation coefficient $\rho$ between $X$ and $Y$ is close
  to 1.
\item The coding rate $R$ is large.
\item The scale factor $s$ is small.
\end{itemize}}
The effect of these assumptions is illustrated in Fig.~\ref{fig:assumptions}.

The basic intuition on which our analysis in this section is built is
very simple: by considering high enough correlations, the encoder can
``roughly center'' the conditional distribution $f_{X|Y}$ at the centroid
of a sublattice cell, a cell that is large enough to make the probability
that the source vector ${\bf x}$ is not in the considered cell negligible,
but at the same time small enough so that tools employed in classical
quantization problems can be applied.

Recall that as mentioned earlier, unlike in classical high rate asymptotics
where $R\to\infty$ results in $\nu(\Lambda)\to 0$, in this case we must
explicitly force $s\to 0$, but not ``too fast'' -- in this case, too fast
would be at a rate equal or faster than the rate at which $f_{X|Y}$ shrinks,
as $|\rho|\to 1$.  We will do so by setting the scale factor $s$ to be
$s = s(\rho)$, where $s:(-1,1)\to\mathbb{R}^+$ is such that 
\begin{eqnarray}
\lim_{|\rho|\to 1} s(\rho) & = & 0, \nonumber \\
\lim_{|\rho|\to 1} \frac{s(\rho)}{\sigma_X\sqrt{1-\rho^2}} & = & \infty.
  \label{eq:choice-s}
\end{eqnarray}
For example,
$s = \sigma_X\sqrt{1-\rho^2}\log\left(1\big/\sigma_X\sqrt{1-\rho^2}\right)$
satisfies these conditions.

\subsubsection{Performance Metric}

Some justification seems necessary at this point for considering
high-correlation asymptotics (i.e., $|\rho|\to 1$), since under this
assumption, the side information available uncoded at the decoder
already contains almost all of the information about the source.  And
indeed, once we are done with our calculations, we will confirm the
(hardly surprising) fact that for any fixed target distortion $D$,
using these proposed quantizers and as $|\rho|\to 1$, the rate required
to achieve $D$ vanishes.  This is a condition that must be satisfied
by {\em any} decent quantizer.  However, that is not why we are
interested in this analysis: instead, our goal is to evaluate
\begin{equation}
  \lim_{|\rho|\to 1} \frac {\bar{d}}{D(R)},
  \label{eq:figure-of-merit}
\end{equation}
where $\bar{d}$ is the distortion of our quantizers, and $D(R)$ is
the Wyner-Ziv rate/distortion function--that is, we wish to compare
the {\em slope} of the distortion function for our proposed quantizers
at asymptotically high correlations, with that of the Wyner-Ziv
bound.  This {\em is} a meaningful performance metric, as it determines
the rate of decay of distortion relative to the fastest possible
decay.\footnote{This type of analysis is similar in spirit to (and
inspired by) that of Verd\'u for modulation schemes operating at
asymptotically low SNRs~\cite{Verdu:02}.}

\subsection{Asymptotics of the Average Error With Geometrically Similar
  Coarse and Fine Lattices}
\label{sec:average-error}

\subsubsection{A Simpler Expression}

To obtain a simpler expression for $\bar d$ than that of
eq.~(\ref{eq:distortion}), we start by expanding it in a different way:
\begin{eqnarray}
\bar d
  & \stackrel{(a)}{=} &
        \mbox{$\frac 1 n$} \int_{{\bf x}\in\mathbb{R}^n} \int_{{\bf y}\in\mathbb{R}^n}
        ||{\bf x} - \gamma({\bf x},{\bf y})||^2 f_{XY}({\bf x},{\bf y})
        \mbox{d}{\bf x}\mbox{d}{\bf y}
        \nonumber \\
  & = & \mbox{$\frac 1 n$}
        \int_{{\bf y}\in\mathbb{R}^n}\left[\int_{{\bf x}\in\mathbb{R}^n}
        ||{\bf x} - \gamma({\bf x},{\bf y})||^2 f_{X|Y}({\bf x}|{\bf y})
        \mbox{d}{\bf x}\right] f_Y({\bf y})\mbox{d}{\bf y}
        \nonumber  \\
  & \stackrel{(b)}{=} & \mbox{$\frac 1 n$}
        \sum_{\lambda\in s\Lambda} \int_{{\bf y}\in V[\lambda:s\Lambda]}
        \left[\int_{{\bf x}\in\mathbb{R}^n}
        ||{\bf x} - \gamma({\bf x},{\bf y})||^2 f_{X|Y}({\bf x}|{\bf y})
        \mbox{d}{\bf x}\right] f_Y({\bf y})\mbox{d}{\bf y}
        \nonumber \\
  & \stackrel{(c)}{\approx} &
        \mbox{$\frac 1 n$} \sum_{\lambda\in s\Lambda}
        \left[ \int_{{\bf x}\in\mathbb{R}^n}
        ||{\bf x}-\gamma({\bf x},\lambda)||^2 f_{X|Y}({\bf x}|\lambda)
        \mbox{d}{\bf x} \right] f_Y(\lambda)\nu(s\Lambda)
        \nonumber \\
  & \stackrel{(d)}{=} &
        \mbox{$\frac 1 n$}
        \left[ \int_{{\bf x}\in\mathbb{R}^n}
        ||{\bf x}-\gamma({\bf x},\mathbf{0})||^2 f_{X|Y}({\bf x}|\mathbf{0})
        \mbox{d}{\bf x} \right]
        \left(\sum_{\lambda\in s\Lambda} f_Y(\lambda)\nu(s\Lambda)\right)
        \nonumber \\
  & \stackrel{(e)}{\approx} & \underbrace{\mbox{$\frac 1 n$}
        \int_{{\bf x}\in V[{\bf 0}:s\kappa(\Lambda)]}
        ||{\bf x}-\gamma_k({\bf x})||^2 f_{X|Y}({\bf x}|\mathbf{0})
        \mbox{d}{\bf x}}_{\alpha}
        \\ & & \mbox{\hspace{2mm}} + \underbrace{\mbox{$\frac 1 n$}
        \sum_{\lambda\in s\kappa(\Lambda)\backslash\{{\bf 0}\}}
        \int_{{\bf x}\in V[\lambda:s\kappa(\Lambda)]}
        ||{\bf x}-\big(\lambda+\gamma_k({\bf x})\big)||^2
        f_{X|Y}({\bf x}|\mathbf{0}) \mbox{d}{\bf x}}_{\beta}
        \label{eq:def-alpha-beta}
\end{eqnarray}
where:
\begin{itemize}
\item[\small $(a)$] is again just the definition of average distortion;
\item[\small $(b)$] follows from partitioning the set of all side information
  sequences ${\bf y}$ into Voronoi cells of the fine lattice $s\Lambda$;
\item[\small $(c)$] follows from the assumption that $\nu(s\Lambda)$ is
  small, and from the continuity of $\int_{{\bf x}\in\mathbb{R}^n}
  ||{\bf x} - \gamma({\bf x},{\bf y})||^2 f_{X|Y}({\bf x}|{\bf y})
  \mbox{d}{\bf x}$ as a function of $\mathbf{y}$;
\item[\small $(d)$] follows from the symmetry of $f_{X|Y}$ as a function
  $\mathbf{y}$;
\item[\small $(e)$] follows from the fact that $f_Y$ integrates to 1, and
  from splitting the domain of integration of ${\bf x}$ into Voronoi cells of
  the sublattice $s\kappa(\Lambda)$.
\end{itemize}
Our next goal is to find simpler expressions for $\alpha$ and $\beta$.

\begin{figure}[ht]
\centerline{\psfig{file=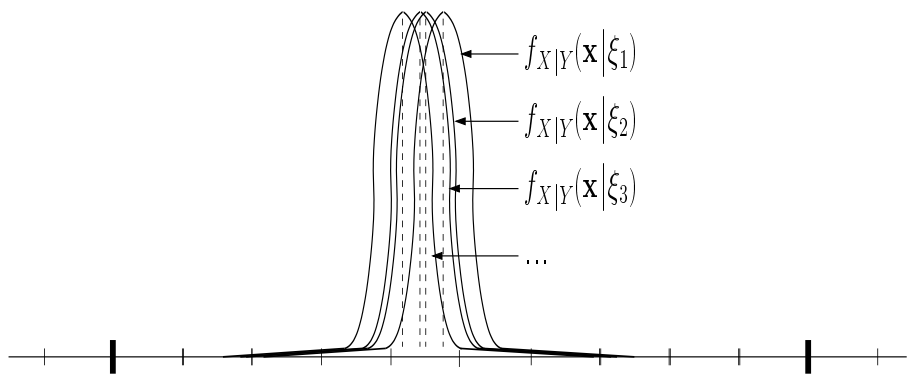,height=5cm,width=12cm}}
\caption{Illustration (in one dimension) of the concept that, irrespective
  of a small random shift in the mean introduced by
  the unknown side information, a fine quantization of the sublattice cell
  (thin lines in between thick lines) results in a fine quantization of
  the unknown distribution.  The true distribution could be any of those
  illustrated for various unknown vectors $\xi_k$.}
\label{fig:simplify-alpha}
\end{figure}

To simplify $\alpha$, we observe that this term denotes the MSE incurred
into when quantizing samples of a distribution $f_{X|Y}({\bf x}|\xi)$
with an $N$-level fixed-rate {\em uniform} quantizer, if we assume that
the overload cells of the quantizer occur with negligible probability --
and this assumption is justified because, for $|\rho|\approx 1$, sublattice
cells are large relative to the spread of $f_{X|Y}$ due to our choice of
$s$ in~(\ref{eq:choice-s}).  Now, again under the assumption that $R$ is
large, the random shift in the mean of $f_{X|Y}$ given by its dependence
on the unknown parameter $\xi$ is negligible compared to the size of a
sublattice cell.  Thus, by choosing a value of $|\rho|$ close enough to
1, the probability of ${\bf x}\not\in V[{\bf 0}:s\kappa(\Lambda)]$ can
be made arbitrarily small.  This is illustrated in
Fig.~\ref{fig:simplify-alpha}.

The requirement that the fine and coarse quantizers be geometrically
similar lattices results in cells of the coarse lattice being partitioned
{\em uniformly} by the fine lattice; this is the optimal quantizer for
a source that is uniformly distributed over a sublattice cell, not
distributed according to $f_{X|Y}$.  Therefore, defining a new pdf
$p(\mathbf{x})=\frac 1{s^nN}$ if $\mathbf{x}$ is in the corresponding
sublattice cell, and zero otherwise, we have that
\[ \lim_{N\to\infty}N^{\frac 2 n}\alpha = G(\Lambda)s^2;
\]
this follows from evaluating eqn.~(81) in~\cite[Ch.\ 2]{neil:splag}
for the uniform distribution $p$ defined above, specialized to the
lattice $\Lambda$.  Therefore, for $N$ large, we can (equivalently)
say that
\[ \alpha \;\;\approx\;\; G(\Lambda)s^2e^{-2R}.
\]

Since $\beta\geq 0$, we have that $\bar d\geq\alpha$, and so
\begin{eqnarray}
\bar d
  & \geq & G(\Lambda)\,s^2\,e^{-2R}.
  \label{eq:distortion-aroundzero-similarcoarsefine}
\end{eqnarray}

\subsubsection{Comparison Against Wyner's Rate/Distortion Bound}

Our next step is to evaluate the figure of merit defined
by~(\ref{eq:figure-of-merit}).  To this end, consider Wyner's
rate/distortion bound~\cite{Wyner:78}:\footnote{In
Wyner's paper, the bound is given in the form $R(d)=\frac{1}{2}\log\left(
\frac{\sigma_X^2\sigma_U^2}{(\sigma_X^2+\sigma_U^2)d}\right)$ (for
the low distortion region), where $\sigma_X^2$ is the variance of $X$,
and $Y=X+U$, where $U$ has variance $\sigma_U^2$.  A straightforward
manipulation puts Wyner's expression in the form shown here.}
\begin{equation}
  D(R) = \sigma_X^2(1-\rho^2)e^{-2R}.
  \label{eq:wynerziv-rdfunction}
\end{equation}
Plugging eqns.~\eqref{eq:distortion-aroundzero-similarcoarsefine}
and~\eqref{eq:wynerziv-rdfunction} into~(\ref{eq:figure-of-merit}), we get
\begin{eqnarray*}
\lim_{|\rho|\to 1} \frac {\bar{d}}{D(R)}
  & \geq & \lim_{|\rho|\to 1}
        \frac{G(\Lambda) s^2 e^{-2R}}
             {\sigma_X^2(1-\rho^2)e^{-2R}} \\
  & = & G(\Lambda)
        \lim_{|\rho|\to 1}\frac{s^2}{\sigma_X^2(1-\rho^2)} \\
  & = & \infty;
\end{eqnarray*}
the divergence of this limit follows from choice of lattice scaling
specified in eqn.~\eqref{eq:choice-s}.  Therefore, when the fine
quantizer is constrained to be a lattice that is geometrically similar
to the coarse lattice, the performance of the resulting Wyner-Ziv
quantizer is very poor in the asymptotic regime of high correlations.
This observation motivates us to introduce a small modification in
our code construction.

\subsection{Asymptotics of the Average Error with a Coarse Lattice and
  an Optimal Fixed-Rate Fine Quantizer}

\subsubsection{A Simpler Expression}

The suboptimality of the code construction based on two geometrically
similar lattices stems from the fact that sublattice cells are partitioned
uniformly, but the source distribution $f_{X|Y}$ being quantized is not
uniform.  Therefore, we enlarge the class of codes considered:
\begin{itemize}
\item we keep the requirement that the coarse quantizer be a lattice;
\item we keep the same quantization algorithm of eqn.~\eqref{eq:q-alg};
\item but we now allow for the fine quantizer to be any arbitrary
  fixed-rate classical vector quantizer.
\end{itemize}
By removing the restriction that the fine quantizer also be a lattice,
we can now choose one still with $N$ reconstruction points, but whose
output point density, instead of being uniform, is matched to the
distribution $f_{X|Y}(\mathbf{x}|\mathbf{0})$.  As a result, we conclude
that there exists a quantizer such that
\[ \lim_{N\to\infty} N^{\frac 2 n}\alpha\;\;=\;\;G_n||f_{X|Y}||_{\frac{n}{n+2}},
\]
where $||f||_{\frac{n}{n+2}} \triangleq \big[ \int f^{\frac{n}{n+2}}(x)
\mbox{d}x \big]^{\frac{n+2}{n}}$, and where $G_n$ depends only on $n$ (but
not on the source distribution), and is bounded in terms of the standard
$\Gamma$ function by
\begin{equation}
   \frac 1{(n+2)\pi}\;\Gamma\Big(\frac n 2+1\Big)^{\frac 2 n}
   \;\;\leq\;\;
   G_n
   \;\;\leq\;\;
   \frac 1{n\pi}\;\Gamma\Big(\frac n 2+1\Big)^{\frac 2 n}
                \;\Gamma\Big(1+\frac 2 n\Big),
   \label{eq:bounds-Gn}
\end{equation}
as follows from eqns.~(81) and~(82) of~\cite[Ch.\ 2]{neil:splag}.
Hence, for $|\rho|\approx 1$ and for $N$ large, we can approximate
$\alpha$ by
\[ \alpha\;\;\approx\;\;G_n\,||f_{X|Y}||_{\frac{n}{n+2}}\,e^{-2R}.
\]

To simplify $\beta$, the following estimate is obtained in
Appendix~\ref{app:trivial1}:
\begin{equation}
  \beta \;\; \approx \;\; \mbox{$\frac 1 n$}
        \frac{2\nu(s\kappa(\Lambda))e_ns^2}
             {[2\pi\sigma_X^2(1-\rho^2)]^{\frac{n}{2}}}
        \left(\frac{e^{-\frac{s^2}{2\sigma_X^2(1-\rho^2)}}}
                   {1-e^{-\frac{s^2}{2\sigma_X^2(1-\rho^2)}}}\right).
  \label{eq:b}
\end{equation}

Combining these two estimates, we arrive at a final expression for
$\bar d$:
\begin{eqnarray}
\bar d
  & \approx & G_n\,||f_{X|Y}||_{\frac{n}{n+2}}\,e^{-2R}
        + \mbox{$\frac 1 n$}
        \frac{2\nu(s\kappa(\Lambda))e_ns^2}
             {[2\pi\sigma_X^2(1-\rho^2)]^{\frac{n}{2}}}
        \left(\frac{e^{-\frac{s^2}{2\sigma_X^2(1-\rho^2)}}}
                   {1-e^{-\frac{s^2}{2\sigma_X^2(1-\rho^2)}}}\right)
        \label{eq:distortion-aroundzero}
\end{eqnarray}

\subsubsection{Comparison Against Wyner's Rate/Distortion Bound}

Plugging eqns.~\eqref{eq:wynerziv-rdfunction}
and~\eqref{eq:distortion-aroundzero} into~(\ref{eq:figure-of-merit}),
we now get
\begin{eqnarray*}
\lim_{|\rho|\to 1} \frac {\bar{d}}{D(R)}
  & = & \lim_{|\rho|\to 1}
        \frac{G_n ||f_{X|Y}||_{\frac{n}{n+2}} e^{-2R}
              + \mbox{$\frac 1 n$}
                \frac{2\nu(s\kappa(\Lambda))e_ns^2}
                     {[2\pi\sigma_X^2(1-\rho^2)]^{\frac{n}{2}}}
                \left(\frac{e^{-\frac{s^2}{2\sigma_X^2(1-\rho^2)}}}
                           {1-e^{-\frac{s^2}{2\sigma_X^2(1-\rho^2)}}}\right)}
             {\sigma_X^2(1-\rho^2)e^{-2R}} \\
  & = & G_n
        \lim_{|\rho|\to 1}\frac{||f_{X|Y}||_{\frac{n}{n+2}}}
                               {\sigma_X^2(1-\rho^2)}
        + \;\; \lim_{|\rho|\to 1} \mbox{$\frac 1 n$}
          \frac{2\nu(s\kappa(\Lambda))e_ns^2}
               {[2\pi\sigma_X^2(1-\rho^2)]^{\frac{n}{2}}}
          \left(\frac{e^{-\frac{s^2}{2\sigma_X^2(1-\rho^2)}}}
                     {1-e^{-\frac{s^2}{2\sigma_X^2(1-\rho^2)}}}\right)
          \frac{1}{\sigma_X^2(1-\rho^2)e^{-2R}}.
\end{eqnarray*}

From eqn.~(57) in~\cite{zador:quantization-asymptotics}, we have that
$\lim_{n\to\infty} ||f_n||_{\frac{n}{n+2}} = e^{2h(f)}$, where $f_n=(f)^n$
is the $n$-dimensional source distribution, and $h$ denotes differential
entropy.  We don't know of a way to simplify this expression for small
$n$, so we approximate it with its limit value as $n$ gets
large.\footnote{It is important to emphasize that although we consider
large blocks to simplify $||f_n||_{\frac{n}{n+2}}$, this does {\em not}
mean that the distortion expression thus obtained is only valid for high
dimensional quantizers: we can consider long source blocks, in which
small sub-blocks are quantized with low dimensional codes (for example,
{\em scalar} quantizers), and this form would still apply.}
For the conditional Gaussian distribution,
$h(f) = \frac 1 2 \log\big(2\pi e\sigma_X^2(1-\rho^2)\big)$, and hence
\[ G_n\lim_{|\rho|\to 1}
  \frac{\lim_{n\to\infty}||f_{X|Y}||_{\frac{n}{n+2}}}
  {\sigma_X^2(1-\rho^2)} \;\;=\;\; G_n\;2\pi e. \]
Note as well that the second term vanishes: for $|\rho|\to 1$,
from~(\ref{eq:choice-s}) we have that $s^2/\big(\sigma_X^2(1-\rho^2)\big)
\to\infty$, and thus this expression is dominated by the vanishing term
$e^{-\frac{s^2}{2\sigma_X^2(1-\rho^2)}}$.
Hence, we conclude that, by explicitly scaling the quantizers with
$s$ satisfying conditions~(\ref{eq:choice-s}),
\[ \lim_{|\rho|\to 1} \frac{\bar{d}}{D(R)} \;\;=\;\; G_n\; 2\pi e. \]

Finally, since for $n$ large the upper and lower bounds on $G_n$
given in eqn.~\eqref{eq:bounds-Gn} coincide and take the value
$\frac 1{2\pi e}$~\cite[pg.\ 58]{neil:splag}, we see that indeed,
as $n\to\infty$, there exist high-dimensional codes for which this
limit can be made arbitrarily close to 1.  {\em Hence, asymptotically
in rate and correlation, our code constructions achieve the Wyner-Ziv
bound.}

\subsection{Some Intuitive Remarks}

\subsubsection{On the Optimality of our Codes, in Hindsight}

Informally, these are the key elements contributing to the optimality
of our codes:
\begin{itemize}
\item The codes are scaled in a way such that, as correlation
  increases, the tails of the conditional distribution $f_{X|Y}$
  outside a cell of the coarse quantizer become increasingly light.
\item At high correlations, our scaling of the codes results
  in the size of cells in the coarse quantizer being small.  But
  at high rates, the size of a cell in the fine quantizer is negligible
  even relative to the small coarse cells.  And the side information
  is, with high probability, ``pinned'' within one of the small fine
  quantizer cells.
\item Because the tails of $f_{X|Y}$ are increasingly light
  as correlation increases, and $f_{X|Y}$ is {\em not} uniform,
  an optimal quantizer for a uniform distribution is mismatched
  to the actual statistics of the data, thus resulting in a severe
  penalty in rate.  However, this penalty can be eliminated entirely
  in a very simple way: only changing the shape of the cells for
  the fine quantizer is enough -- if the output point density of
  the fine quantizer is matched to the pinned form of $f_{X|Y}$,
  this is an optimal code.
\end{itemize}
Essentially, our construction is asymptotically optimal (in rate
and correlation), because we scale the lattice in a way such that
we create multiple copies of $f_{X|Y}$ one within each cell of
the coarse lattice, and we use an optimal code within that cell.

\subsubsection{On Why $R^*(d)=R_{X|Y}(d)$ for Gaussian Sources}

This asymptotic analysis also sheds light on why there is no
rate loss for Wyner-Ziv coding of Gaussian sources, at least in
the asymptotic regime of high rates and high correlations.  Note
that the conditional distribution $f_{X|Y}$ depends on the side
information $\mathbf{y}$ only in the form of a random shift: this
random shift becomes negligible at high rates, but more importantly,
the {\em shape} of $f_{X|Y}$ is independent of $\mathbf{y}$.  As
a result, a single code can be used to quantize the $f_{X|Y}$'s
pinned one within each cell of the coarse lattice.  It is this
invariance property of the conditional Gaussian distribution that
results having $R^*(d)=R_{X|Y}(d)$, at least in the asymptotic
regime considered in this section.

\section{Applications in Sensor Networks}
\label{sec:sensor-networks}

\subsection{Discussion}

Issues in the analysis of performance of wireless networks have received
considerable attention in recent times.  To a large extent, interest on
these topics has been sparked by an observation made by Gupta and Kumar:
the total throughput that can be carried by one particular class of
wireless networks is only $O(\sqrt{n})$,\footnote{A word on notation.
In this section, $n$ denotes number of nodes in the network, and $N$
denotes block length.  This notation should not be confused with that
in previous section, where $n$ was used to refer to block length, and
$N$ to the number of reconstruction codewords in a code.}
for a network having $O(n)$ nodes~\cite{GuptaK:00}.  As a result, each
source-destination pair gets a throughput of $O(1/\sqrt{n})$, i.e., the
amount of information that any one individual node can inject into the
network vanishes as the network size increases.  The model used for
performance analysis in~\cite{GuptaK:00} was conceived as an abstraction
for emerging ad-hoc wireless networks, made up of small appliances (such
as laptop computers or microwave ovens or door locks), interconnected
via standard air interfaces (such as Bluetooth or 802.11).  In that
context, the fact that as more nodes join the network then the capacity
available to each node decreases, clearly poses serious problems, since
there is no reason to believe that there will be any dependencies in the
data generated by each of these devices.  And these problems prompted
the conclusion in~\cite{GuptaK:00} that networks with either a small
number of nodes, or with a small number of connections, may be more
likely to find acceptance.

In our work, we consider a different type of wireless networks: we
focus on {\em sensor} networks, i.e., networks of devices that collect
measurements of a process that is ``regular'' in some sense.  For example,
if the sensors measure ozone concentration in the atmosphere, then the
values of each measurement will not be independent in general, but instead
will be constrained by an appropriate form of the Navier-Stokes equations.
If the sensors measure temperatures at different locations of a material,
the measurements will be constrained by Fourier's heat equations.  And
in general, when the sensors sample values of some random process at
different locations, these samples will be constrained by the correlation
structure of the process (see, e.g.,~\cite{ServettoR:06}).  By considering
correlated sources we generalize in what we believe is a very meaningful
way the setup of~\cite{GuptaK:00}: now the amount of information generated
by each node is no longer a constant, but instead it depends on the size
of the network itself.

\subsection{Network Model}

Consider the following problem setup:

\begin{itemize}
\item There is a source of information, modeled by a process $X_u(k)$: for
  fixed values of $k$, $X_u(k)$ is a brownian motion with parameter
  $\sigma^2$; for fixed values of $u\in[0,1]$, $X_u(k)$ is an iid sequence.
  That is, at a fixed location $u$, iid samples with distribution
  $N(0,\sigma^2u)$ are collected in discrete time, and at a fixed time
  slot, a Wiener process unfolds in space.
\item Network nodes are represented by points on the unit square
  $[0,1]\times[0,1] \subset \mathbb{R}^2$, and are classified into
  three groups:
  \begin{itemize}
  \item There are $n$ {\em source} nodes $s$, that feed information into
    the network, uniformly spread on the left edge of the square.
  \item There are $n$ {\em destination} nodes $d$, that take information
    out of the network, uniformly spread on the right edge of the square.
  \item There are $n$ {\em router} nodes $r$, optimally placed in
    the interior of the square, to maximize network throughput.  These nodes
    are pure routers, they neither inject nor extract information to/from
    the network, and they don't apply any form of coding, they only forward
    information to other nodes.
  \end{itemize}
\item The $m$-th source collects samples of $X_{m/n}(k)$, and
  encodes this information prior to sending it to the $m$-th destination
  ($m=1...n$).  The only information available to each source is:
  \begin{itemize}
  \item The observed samples $X_{m/n}(k)$.
  \item The position in the square of all the nodes.
  \item The statistics of the entire process $X$.
  \end{itemize}
\item Each destination node forwards whatever data it receives to a special
  node $d$, which {\em jointly} decodes all the data received, and computes
  an estimate $\hat{X}_u(k)$ of the entire sample path $X_u(k)$ based on all
  the decoded samples $X_{m/n}(k)$'s.
\item Nodes do not move, and have an unbounded power supply.
\item A bit is successfully sent from node $v_i$ to node $v_j$ if
  (a) $||v_i-v_j||<\Delta_i$, and (b) if for all other transmitting nodes
  $v_k$, $||v_k-v_j|| \geq \Delta_k$.  $R$ bits per channel use can be
  transmitted over any link.
\item Routing and power control are optimally configured to maximize network
  throughput.
\end{itemize}
Note that in this model we explicitly rule out the possibility of
source nodes exchanging information to cooperate in the encoding of their
observations.  Note also that routers only forward data, but do not apply
any form of coding.  That is, encoding is distributed among the sensors,
data is carried over the network by relay nodes, and decoding is
performed at a central location.

We should point out that our model is different from the model of Gupta
and Kumar~\cite{GuptaK:00}: whereas in their model they consider $n$ nodes
which serve as transmitters/receivers/relays all in a single device, we
break up each device into three pieces, and consider $n$ transmitters, $n$
receivers, and $n$ relays.  However, this is not a fundamental difference:
as long as we keep the same number of all three types of devices,
the two models are essentially the same, and therefore their results on
the property of vanishing throughputs as $n\rightarrow\infty$ still holds
for our model.  The idea of splitting the devices into three separate units
is to model a situation in which data is captured at some location, is
transported over an ad-hoc network, and an estimate of the field of
measurements is formed at a remote location.

\subsection{Encoding/Decoding Mechanics in Large Networks}

Clearly, a network with a finite number of nodes and with communication
links of finite capacity among nodes, can transport only a finite amount
of information.  Therefore, exact reconstruction of the brownian field
$X_u(k)$ will not be possible in general, and a key issue then is that
of understanding the rate/distortion tradeoffs involved.  A thorough study
of this new rate/distortion problem lies outside the scope intended for
this paper, and we will deal with this problem elsewhere.  Of interest
in this paper however is a result that relates the ability of the central
destination node $d$ to estimate the brownian field $X_u(k)$ to both the
number of nodes in the network and the capacity of the individual network
links.  Indeed, we have that under the assumption of a large (but still
independent of network size) link capacity $R$, for any $\epsilon>0$ and
$1-\epsilon\leq\rho<1$, there exists a large enough network of size $n$
nodes, such that
\[
    D_{\frac m n} \;\stackrel{\Delta}{=}\;
    {\tt E}\left(||X_{\frac m n}(k)-\hat{X}_{\frac m n}(k)||^2\right) \;\leq\;
      \sigma_X^2\mbox{\small$\frac{m-1}{n}$}(1\!-\!\rho^2)
                    \;e^{-\frac{R}{6\sqrt{n}}}
      \mbox{ (a.e.)},
\]
uniformly for $\frac m n$ in the closed interval
$\left[\frac{1}{n(1-\rho^2)},1\!\right]$, where $m\leq n$ is an integer,
for all time slots $k$, and for almost all sample paths of the field
$X_{\frac m n}(k)$.

Essentially, what this result states is that, under the assumption of a
large network and with links of high capacity, it is possible for $d$
to estimate the sample paths of $X$ with arbitrarily small error.  That
accurate estimation is possible is indeed surprising to us, given the
fact that the amount of information per sample that the network can
carry vanishes~\cite{GuptaK:00}---fortunately, so does the information
content per sample, and that is what we can take advantage of.

\subsubsection{Placement of Nodes and Scheduling of Transmissions}
\label{sec:placement-scheduling}

First of all, we give one particular distribution of routers in the
plane and one particular algorithm for scheduling transmissions.

Assume $\ell = \sqrt{n}$ is an even integer, and define:
\begin{itemize}
\item The sources are located at coordinates $(0,\frac{i}{n})$, and the
  destinations at coordinates $(1,\frac{i}{n})$, for $i=1...n$.
\item There are exactly $n$ routers, located at coordinates
  $(\frac{1}{2\ell}+\frac{i}{\ell},\frac{1}{2\ell}+\frac{j}{\ell})$,
  for $i,j=0,1,...,\ell-1$.
\item The transmission radius for the source nodes is
  $\Delta=\frac{\sqrt{2}}{2\ell}$, and for the routers it is
  $\Delta=\frac{1}{\ell}$.\footnote{Recall that destination nodes do not
  communicate over the shared wireless medium with the central decoder,
  they only receive data that way.  Therefore, no transmission range
  needs be specified in their case.}
\end{itemize}

In order to present an algorithm to schedule transmissions over time,
we need some definitions.  First, divide the square
$[0,1]\times[0,1]\subset\mathbb{R}^2$ into $\ell$ sets defined by
\[
   S^{(i)} = \left[\frac{(i\!-\!1)\ell}{n},\frac{i\ell}{n}\right)\times[0,1]
\]
$(i\!=\!1...\ell)$.  Within each $S^{(i)}$, there are:
\begin{itemize}
\item $\ell$ source nodes, at coordinates
  $\left(0,\frac{(i-1)\ell+m}{n}\right)$, for $m=0...\ell-1$.
\item $\ell$ destination nodes, at coordinates
  $\left(1,\frac{(i-1)\ell+m}{n}\right)$, for $m=0...\ell-1$.
\item $\ell$ router nodes, at coordinates
  $\left(\frac{1}{2\ell}+\frac{k-1}{\ell},
         \frac{1}{2\ell}+\frac{i}{\ell}\right)$, for $k=1...\ell$.
\end{itemize}
Next, we divide the router nodes into three groups $g_0,g_1,g_2$: a
router falls in $g_j$ if its index $k$ is equal to $j$ (mod 3).  Source
nodes all belong to the group $g_0$.  Finally, we give an algorithm to
schedule transmissions:
\begin{itemize}
\item Time is discrete, and starts at 0.  At even time slots, allow
  transmissions of nodes in $S^{(i)}$'s for which $i$ is even; at odd time
  slots, allow transmissions of nodes for odd $i$'s.
\item Each $S^{(i)}$ keeps its own clock $\tau_i$, which advances only
  when transmissions from this $S^{(i)}$ are allowed to proceed: when
  $\tau_i\equiv 0$ (mod 3) then $g_0$ sends, when $\tau_i\equiv 1$ (mod 3)
  then $g_1$ sends, when $\tau_i\equiv 2$ (mod 3) then $g_2$ sends.  And
  source nodes send only once every $\ell$ available slots, cycling through
  them in round-robin order.
\end{itemize}

\begin{figure}[!h]
\vspace{-3mm}
\centerline{\psfig{file=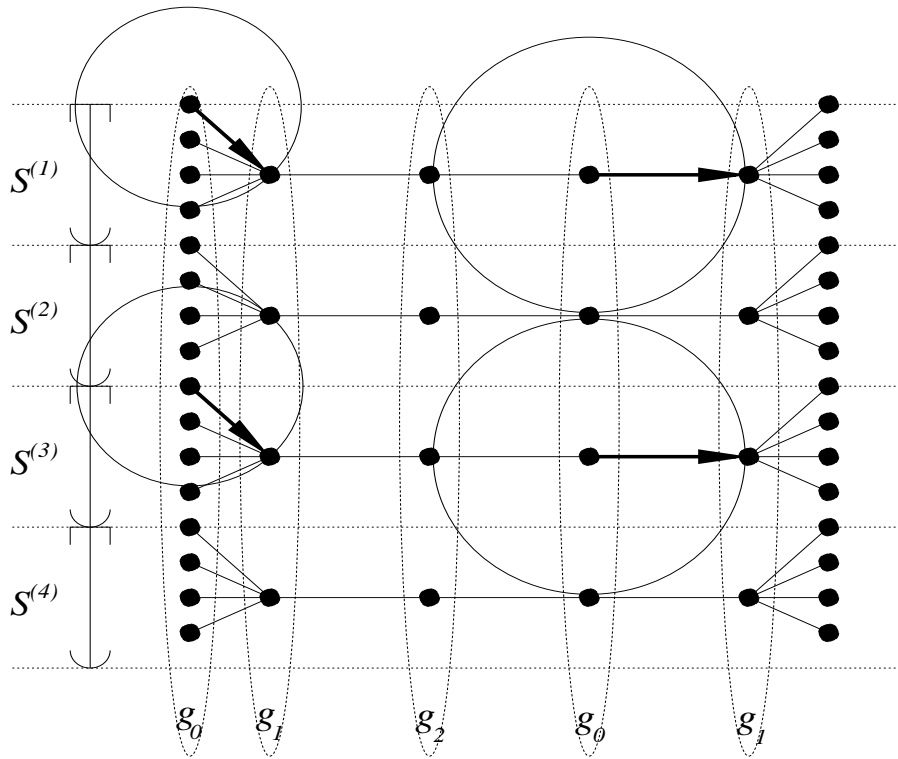,height=10cm,width=12cm}}
\vspace{-3mm}
\caption{An example of the placement and division of nodes, and
  scheduling of transmissions, for $n=16$ ($\ell=4$).  Black dots represent
  nodes: 16 sources on the left edge of the square, 16 routers inside the
  square, 16 destinations on the right edge of the square.  A source sends
  data to a destination on the same horizontal line.  Thin solid lines
  joining nodes are
  routes.  The sets $S^{(i)}$ and the groups $g_i$ are indicated with dotted
  lines.  Active transmissions are indicated with a thick arrow, and the
  circles around each indicate transmission ranges.  The active
  transmissions in this picture correspond to an odd time slot (nodes only
  within $S^{(1)}$ and $S^{(3)}$ are sending), and the group $g_0$ is active.}
\label{fig:step1}
\end{figure}

An illustration of the placement and divisions of nodes, and of the
mechanics of the algorithm, is shown in Fig.~\ref{fig:step1}.

\subsubsection{Throughput per-Node is $\frac{R}{6\sqrt{n}}$}

The calculation of throughput proceeds in three steps:
\begin{enumerate}
\item Each group $S^{(i)}$ is scheduled for transmission only $\frac{1}{2}$
  of the available time slots.  Among these slots, only $\frac{1}{3}$ are
  available for transmission by $g_0$, the group that contains source nodes.
  When this group is scheduled, only once every $\ell$ slots is available
  to a particular node.  And when a particular node finally gets his
  chance to inject a message into the network, it injects $R$ bits (equal
  to link capacity).  Therefore, the total number of bits {\em injected} by
  any one source node per unit of time is
  $\frac{1}{2}\frac{1}{3}\frac{1}{\ell}R=\frac{R}{6\sqrt{n}}$.
\item By construction, there is never more than one packet of $R$ bits in
  the buffer of any router.
\item Also by construction, there is never more than one active transmission
  within range of any receiver.
\end{enumerate}
So, from 1 we have that $\frac{R}{6\sqrt{n}}$ bits per time slot are
injected into the network, from 2 we have that there is no buildup of
packets in any one queue, and from 3 we have that packets are never lost
or delayed.  Therefore, all injected bits reach destination, and hence
the throughput is $\frac{R}{6\sqrt{n}}$ bits per time slot per node.

\subsubsection{Use of Codes with Side Information}
\label{sec:use-lqsi}

So far we have a network in which there is no loss of data, and which
can carry a total of $\frac{R}{6\sqrt{n}}$ bits per time slot per node.
And we collect one sample of the brownian field $X$ per time slot at
each source node.  Therefore, we have $\frac{R}{6\sqrt{n}}$ bits per
sample to encode a block of $N$ samples, for which the network guarantees
delivery.

Consider encoding a block of samples $X_{m/n}^N
\stackrel{\Delta}{=} [X_{m/n}(0)...X_{m/n}(N-1)]$ at the $m$-th
source node.  Trivially, we have that
$X_{m/n}^N = X_{(m-1)/n}^N +
(X_{m/n}^N-X_{(m-1)/n}^N)$.  From standard properties of
Wiener processes, we have that $X_{m/n}^N$ and $X_{(m-1)/n}^N$
are jointly Gaussian, and that the increment has distribution
\[
   X_{m/n}^N-X_{(m-1)/n}^N
     \;\sim\; N\left(0,\mbox{$\frac{\sigma_X^2}{n}$}{\bf I}\right),
\]
independent of $X_{(m-1)/n}^N$.  If $X_{(m-1)/n}^N$ were
available at the $m$-th encoder, the encoding procedure would be trivial:
use standard codes for an iid Gaussian source to send this increment.  But
without the reference value $X_{(m-1)/n}^N$, $m$ cannot compute that
increment, which is the only ``new'' information at location
$\frac{m}{n}$.

Our encoding procedure is as follows: we encode $X_{m/n}^N$ using the
codes developed in earlier sections, assuming the side information
$X_{(m-1)/n}^N$ is available at the decoder.  The relevant statistics
are:
\[
   X_{(m-1)/n}^N
     \sim N\left(0,\sigma_X^2(m\!-\!1)/n{\bf I}\right),\hspace{8mm}
   X_{m/n}^N \sim N\left(0,\sigma_X^2m/n{\bf I}\right),\hspace{8mm}
   \rho_{m-1,m} = \sqrt{1-1/m}.
\]

\subsection{Distortion Computation}

Next we turn to the computation of distortion for this proposed coding
strategy.  Note that since the side information used to decode the data
generated by one node is the data available at previous nodes, and that
decoding errors can indeed occur with non zero probability (and thus,
in the large-network regime, {\em will} occur), an important issue that
needs to be addressed is the effect of decoding errors on the overall
achieved distortion.

We proceed in two steps: first we compute the distortion resulting
in the case when no decoding errors occur, and then we compute the increase
in distortion due to decoding errors.

\subsubsection{Distortion Assuming No Decoding Errors}

Consider a fixed location $\frac m n$ ($1\leq m\leq n$), a fixed
desired correlation value $\rho$ based on which a large enough value
of $n$ is determined, and assume that no decoding errors occur in
decoding samples $\frac 1 n ... \frac{m-1}n$.

\begin{figure}[!ht]
\centerline{\psfig{file=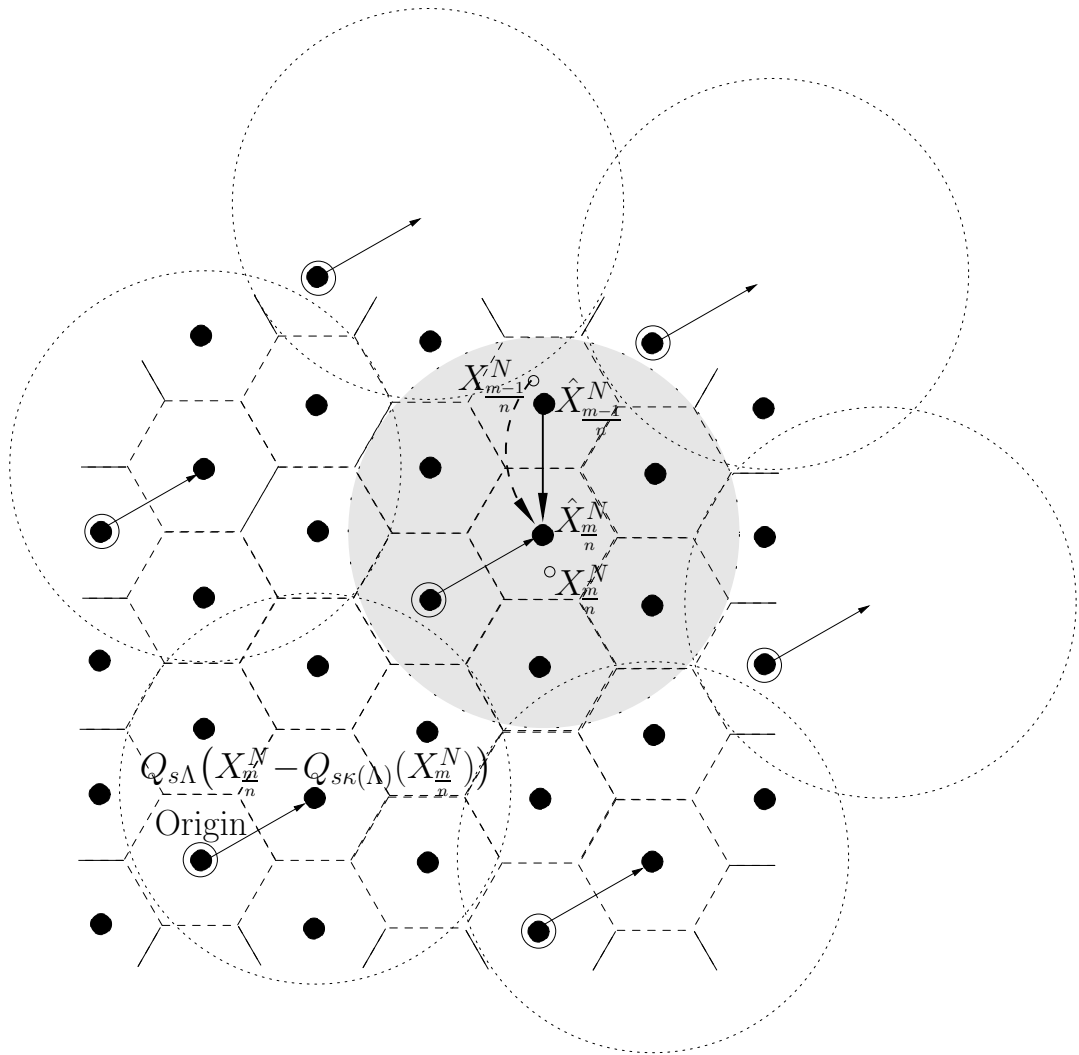,height=14cm}}
\caption{To illustrate the robustness of the proposed quantizers
  to small amounts of quantization noise in the side information: as long
  as the side information falls within a sublattice cell (roughly indicated
  as the shaded region in this picture), using coded or uncoded side
  information does not make a difference.  In this case, $X^N_{\!\frac{m-1}n}$
  is the sample at the previous location, used as side information for the
  sample $X^N_{\!\frac m n}$ at the current location.}
\label{fig:coded-sideinfo}
\end{figure}

In Section~\ref{sec:use-lqsi} above, we argued that we can use
codes with side information to effectively approximate the performance
of a genie-aided encoder capable of sending the increments at each node.
We would like to point out now that in our decoder, the side information
is itself quantized with the coarse lattice.  As a result, as long as
$X_\frac{m-1}n$ and $\hat X_\frac{m-1}n$ fall in the same sublattice
cell, the reconstruction $\hat X_\frac m n$ is as good as if it were
based on {\em uncoded} side information.  This is illustrated in
Fig.~\ref{fig:coded-sideinfo}.

Thus we conclude that, provided no decoding errors occur in any of the
previous samples, and based on the results in Section~\ref{sec:asymptotics},
we can approximate the distortion in the reproduction of each sample
by Wyner's rate/distortion bound:
\[
   D_\frac{m}{n} \:\leq\:
    \sigma_X^2\mbox{$\frac{m}{n}$}(1\!-\!\rho^2)\;e^{-\frac{R}{6\sqrt{n}}},
\]
Note that the inequality in this case is because there will be nodes
operating with a correlation value higher than the specified $\rho$, and
for these values $D_u$ will be even lower than this.  The location-dependent
correlation coefficients $\rho_{m-1,m}$ between adjacent samples forms a
monotonically increasing sequence $\sqrt{1-1/m}\longrightarrow 1$ as
$m\rightarrow\infty$.  A trivial manipulation shows that for all
$m\geq\frac{1}{1-\rho^2}$, $\rho\leq\rho_{m-1,m}<1$, and therefore all node
locations $\frac{m}{n}$ in the closed interval $\left[\frac{1}{n(1-\rho^2)},
1\right]$ will have correlation values at least $\rho$.  Now, since
$m\leq n$, by choosing $n$ large enough we can make $\frac{1}{n(1-\rho^2)}$
come arbitrarily close to zero.  So we see that the distortion bound above
holds uniformly for almost all samples in a large network.

At locations $u$ in which there is no sample collected (i.e., any location
in an open interval $\left(\frac{m-1}{n},\frac{m}{n}\right)$), we need to
interpolate $X_u$: we define $\hat{X}_u = \hat{X}_{(m-1)/n}$, where
$(m-1)/n<u<m/n$.\footnote{Note that we could use better interpolators here
than a simple zero-order hold.  But already with this rather simple minded
rule we get the sought result of vanishing estimation error, and hence we
keep it for simplicity.}  In this case,
\[
   D_u \leq D_\frac{m-1}{n} + \mbox{$\frac{\sigma_X^2}{n}$},
\]
since the interpolation error is at most the size of an increment
between samples, and this increment has variance $\sigma_X^2/n$.  Assume
now that the sample path $X_u(k)$ is continuous at $u$:
\begin{itemize}
\item Because $n$ is large, and for a fixed $k\in\mathbb{N}$, we have
  a dense sampling of $X_u(k)$, $0\leq u\leq 1$.
\item Because $R$ is large, encoded samples $\hat{X}_u$ available at
  the decoder are close to the original value $X_u$, i.e.,
  $\hat{X}_u\rightarrow X_u$, $u=\frac{m}{n}$.
\item Because $X_u$ is continuous and $n$ is large, we have that
  interpolated samples $X_u\approx X_{(m-1)/n}$
  ($\frac{m-1}{n}<u<\frac{m}{n}$), for all $0\leq u\leq 1$.
\end{itemize}
Therefore, $D_u \leq D_\frac{m-1}{n} + \frac{\sigma_X^2}{n}$ holds at
all points of continuity of $X_u$.  But finally, since almost all paths
of a Wiener process are continuous~\cite{StarkW:94}, we conclude that
\[
   D_u \;\leq\;
   \sigma_X^2\left(\mbox{$\frac{m-1}{n}$}(1\!-\!\rho^2)
   \;e^{-\frac{R}{6\sqrt{n}}}+\mbox{$\frac{1}{n}$}\right)
   \;\; \mbox{(a.e.),}
\]
where $(m-1)/n<u<m/n$, and $1\leq m\leq n$.

\subsubsection{Distortion Excess Due to Decoding Errors}

In the subsection above we obtained an expression for the distortion
in the reconstruction of the sample paths assuming that decoding errors
never occur.  This is clearly a lower bound on the achievable distortion.
But we still need to account for the distortion increase that results
from the increasingly likely (as $n\to\infty$) event of a decoding error.
Our next goal is to show that, in large networks, this excess distortion
is negligible compared to the distortion above induced by the quantizers.

\begin{figure}[ht]
\centerline{\psfig{file=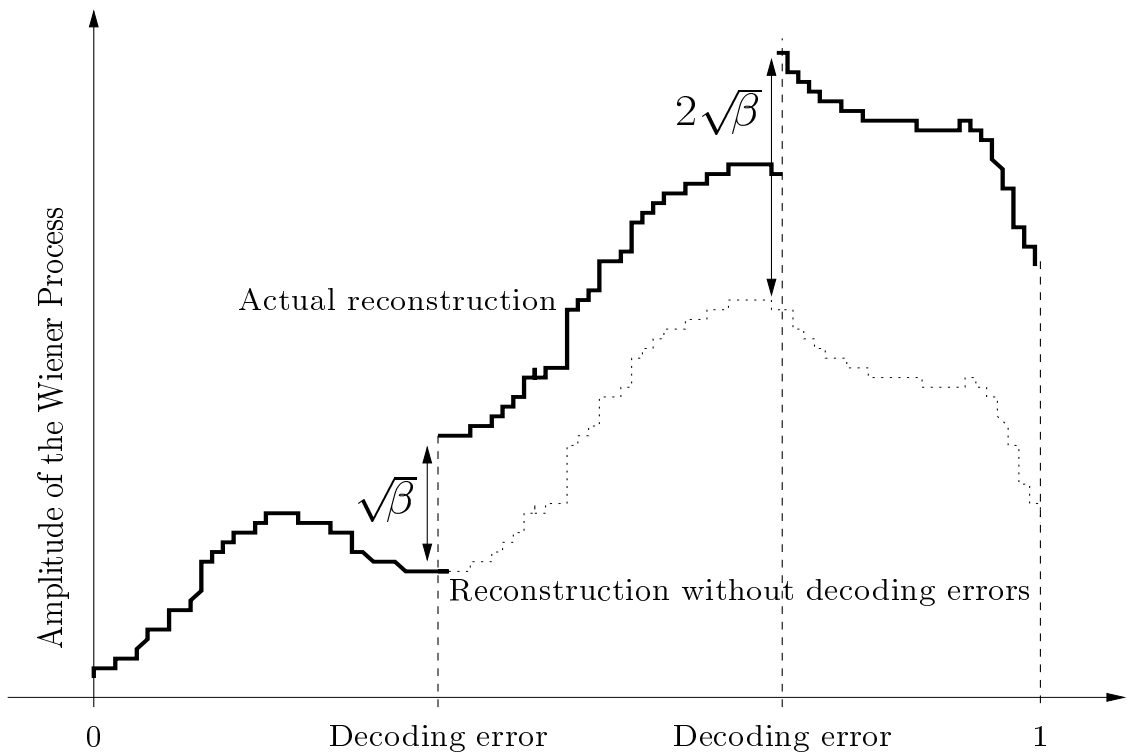,width=15cm,height=10cm}}
\caption{To illustrate the concept of excess distortion.  In this picture
  we show the reconstruction that would result when no decoding errors
  occur (bottom sample path), and the effects of decoding errors (jumps
  of average size $\sqrt{\beta}$, as defined in eqn.~(\ref{eq:def-alpha-beta}),
  after each decoding error).  Note that these errors do not necessarily
  add up coherently from node to node, as illustrated in this picture --
  however, taking them to behave in this way provides a valid upper bound
  on the total excess distortion they induce.}
\label{fig:excess-distortion}
\end{figure}

Consider two definitions:
\begin{itemize}
\item $\Upsilon_m$ is a random variable such that $\Upsilon_m = l$ denotes
  the event in which $l$ nodes (out of the $m$ right before the node at
  location $\frac m n$) make a decoding error.  Since conditioned on the
  side information being correct, errors are independent at each node,
  $\Upsilon_m \sim \mbox{B}(m,p_n)$: a binomial distribution with parameters
  $m =$ number of previous nodes, and $p_n = $ probability of decoding
  error given that there are $n$ nodes in the network.
\item We refer to the term $\beta$ defined by eqn.~(\ref{eq:def-alpha-beta})
  as the {\em excess distortion} at node $m$.
\end{itemize}
Both these definitions are illustrated in Fig.~\ref{fig:excess-distortion}.

Consider now the distortion in a reconstruction of $X_{\frac m n}$ based
on coded side information:
\begin{eqnarray*}
E\big(||X_{\frac m n}-\hat{X}_{\frac m n}||^2\big)
  & \stackrel{(a)}{\approx} &
    \alpha_n+\sum_{l=0}^m P(\Upsilon_m = l) \left(l\sqrt{\beta_n}\right)^2 \\
  & = & \alpha_n+\beta_n E(\Upsilon_m^2)
  \;\; = \;\;
        \alpha_n+\beta_n \big(\mbox{Var}(\Upsilon_m)+E^2(\Upsilon_m)\big) \\
  & \stackrel{(b)}{=} & \alpha_n+\beta_n\big(m p_n (1-p_n) + m^2 p_n^2\big)
  \;\; = \;\; \alpha_n+\beta_n m p_n (1+(m-1)p_n) \\
  & \stackrel{(c)}{\leq} & \alpha_n+\beta_n n p_n(1+np_n)
  \;\; \approx \;\; \alpha_n + \beta_n n^2 p_n^2 \\
  & \stackrel{(d)}{\approx} & \alpha_n + e^{-\frac{n}{2\sigma_X^2}} n^2 p_n^2 \\
  & \stackrel{\Delta}{=} & \alpha_n + \beta'_n
\end{eqnarray*}
where:
\begin{itemize}
\item[(a)] follows from eqn.~(\ref{eq:def-alpha-beta}), and from the fact
  that if $l$ errors occured before the decoding of the $m$-th sample, on
  average each error contributes distortion $\beta_n$ and in the worst of
  cases all these errors add up coherently (the dependence of $\alpha$ and
  $\beta$ in eqn.~(\ref{eq:def-alpha-beta}) on $n$ is highlighted by adding
  the subscript);
\item[(b)] follows from the binomial distribution of $\Upsilon_m$;
\item[(c)] follows from the fact that the expression above must hold for
  all $1\leq m\leq n$;
\item[(d)] follows from the fact that for $n$ large, we can neglect the
  polynomial terms associated with the negative exponential, and from the
  fact that $\rho = \sqrt{1-\frac 1 n}$.
\end{itemize}
Clearly, as $n\to\infty$, both $\alpha_n\to 0$ and $\beta'_n\to 0$.
But again, this is not an interesting observation.  The interesting
observation in this case is that still in the presence of coded side
information and decoding errors, in the regime of high correlations,
$\beta'_n$ is negligible compared to $\alpha_n$, and
$E\big(||X_{\frac m n}-\hat{X}_{\frac m n}||^2\big)\approx\alpha_n$:
\[\begin{array}{ccccccccc}
\lim_{n\to\infty} \frac{\alpha_n+\beta'_n}{\alpha_n}
  & = & 1 + \lim_{n\to\infty} \frac{\beta'_n}{\alpha_n}
  & \leq & 1 + \lim_{n\to\infty}
               \frac{\beta'_n}{\sigma_X^2\mbox{$\frac 1 n$}}
  & \leq & 1 + \lim_{n\to\infty}
               \frac{e^{-\frac{n}{2\sigma_X^2}} n^3 p_n^2}{\sigma_X^2}
  & < & 1+\epsilon,
\end{array}\]
for any $\epsilon>0$ and $n$ large enough.  But we also have
$\frac{\alpha_n+\beta'_n}{\alpha_n}>1$ (since $\beta'_n>0$).  Thus,
the excess distortion due to the use of coded side information and
possible decoding errors is negligible compared to the distortion
induced by the quantizers themselves.

To conclude this section, we would like to point out that there is an
interesting tradeoff in this analysis, that works out favorably for us.
Note that by increasing the number of nodes, we increase the number of
places at which errors can occur, and therefore the probability that
some node will make a decoding error is increased.  However, as the
number of nodes increases, the correlation between their measurements
increases as well, and therefore the size of errors is reduced.  And
as the previous analysis shows, a linear increase in the number of nodes
results in an exponential decrease in the size of each error -- hence,
error propagation is {\em not} a problem in this setup.

\section{Conclusions}
\label{sec:conclusions}

In this paper we presented our work on the design and performance
analysis of codes for the problem of rate distortion with side
information, and on the application of those codes in the context
of a problem of data compression for sensor networks.  First, we
gave concrete constructions for the nested codes studied by
Shamai/Verd\'u/Zamir in~\cite{shamai-verdu-zamir:systematic-lossy-coding,
zamir-shamai:almost-there}, effectively answering an open question
raised in~\cite{zamir-shamai:almost-there}.  Then we studied the
distortion performance of our codes, under the assumption of high
correlation between the source and the side information and of
high coding rates: there we showed that our codes attain the
theoretically optimal distortion decay established by Wyner and
Ziv~\cite{Wyner:78, WynerZ:76}.  Finally we computed an upper bound
on the error made in estimating a brownian field based on measurements
collected by very ``cheap'' devices and delivered over a wireless
network.  In this case, even though the per-node throughput of the
network vanishes as its size increases, and even if the nodes are
not allowed to exchange any information at all, we showed how
arbitrarily accurate estimation of the remote field is possible.
To conclude the paper, we would like to comment on some issues that
follow from our work.

Concerning the problem of source estimation, in the presence of constraints
on the available data imposed by the wireless network:

\begin{itemize}

\item The Brownian model for the source considered in this work is
  probably one of the worst cases we could have considered, in the sense
  that the regularity conditions satisfied by this process are minimal.
  For example, almost all of its sample paths are indeed continuous at
  almost all points (something we did use in our analysis); but at the
  same time, almost all sample paths are {\em not} differentiable at almost
  all points.  Furthermore, the crucial assumption of high-resolution
  quantization that enabled us to apply our codes in the presence of
  {\em coded} side information cannot be justified for processes with
  increments of variance $O(n^{-1+\epsilon})$, for any
  $\epsilon>0$---compare this to the $O(n^{-1})$ variance of the increments
  of the model we considered.

\item Interesting questions arise if we consider processes more regular
  than Brownian motion: consider for example the case when $X_u$ is a
  bandlimited signal (since $X_u$ is compactly supported, take its periodic
  extension).  If the samples $X_{m/n}$ were available at the decoder
  without distortion, it follows from Shannon's sampling theorem that
  a network of finite size is enough to achieve a reconstruction with
  zero distortion.  However, this would require network links of infinite
  capacity.  For any finite value of $R$, there are tradeoffs to explore
  between the number of nodes in the network (i.e., the sampling rate) and
  the capacity of the network links (i.e., the accuracy in the representation
  of each sample), since economic constraints may favor one or the other
  option.  This problem has received considerable attention in the signal
  processing and harmonic analysis literature~\cite{CvetkovicV:98, FuchsD:00,
  GoyalVT:98, KrimTMD:99, ThaoV:94}.

\end{itemize}

Concerning coding/quantization.  Whereas our asymptotic analysis was
performed only for jointly Gaussian sources and MSE distortion, it would
be interesting to learn something about the performance of the proposed
quantizers for sources with non-Gaussian statistics and/or other
distortion measures.  An interesting result of Zamir states that,
although the gap between $R_X(d)$ and $R_{X|Y}(d)$ can be unbound, the
gap between the Wyner-Ziv rate/distortion function $R^*_X(d)$ and
$R_{X|Y}(d)$ is bounded, and actually quite small in some cases: 0.5
bits/sample for arbitrary source statistics and MSE distortion, and 0.22
bits/sample for a binary source with Hamming distortion~\cite{Zamir:96}.
In our opinion this is an interesting issue because, should a result
similar to Zamir's hold for the performance of our codes, this would
immediately allow us to conclude that arbitrarily accurate estimation
is possible not just for jointly Gaussian sources, but for any source
statistics.  And even if we do not have a formal proof, it certainly
seems plausible to us that this may be so.

Concerning the type of asymptotics developed in this work.  Tools
employed for theoretical performance analysis in source coding problems
can be roughly classified into two main groups:
\begin{itemize}
\item Large-block asymptotics, as pioneered by
  Shannon~\cite{Shannon:59}.
\item High-rate asymptotics, as pioneered by Zador, Gersho and
  others~\cite{gersho:quantization-asymptotics,zador:quantization-asymptotics}.
\end{itemize}
The asymptotics we considered in this work are of neither type -- instead,
we focused on {\em high-correlation} asymptotics.  And we believe this
type of analysis is one particularly well suited for a new class of source
coding problems, that originate in the context of sensor networks.  This
paper presents one such analysis for a simple toy problem involving a
Brownian process.  More of our work along these lines can be found
in~\cite{LilisZS:04, ScaglioneS:03, ServettoR:06}.

To conclude, we would like to comment on the nature of our contributions
in this paper.  Since the seminal work of Gupta and Kumar~\cite{GuptaK:00},
most of the theory work on wireless networks appears to have been driven
by a desire to find ways to understand, and if possible circumvent, the
fact that the per-node throughput of the network vanishes as the number
of nodes grows.  Implicit in previous work seems to have been present an
assumption that each node has a constant amount of information to transmit,
irrespective of the network size: in this case, the fact that the throughput
per node decreases as the network size increases does indeed pose serious
problems.  However, we feel the asymptotic analysis of~\cite{GuptaK:00} is
better suited to ``networks of small sensors'' than to ``networks of laptop
computers'': whereas there are only so many laptops that one may want to
have in a single room, much higher densities of small sensing nodes are
conceivable.  Yet it is very high densities of nodes what the asymptotic
analysis of~\cite{GuptaK:00} suggests to us.  Now, in the context of sensor
networks, the vanishing-throughput property of some wireless networks is
much less of a problem.  As an application for our codes with side
information, we illustrated an instance of a class of wireless networking
problems in which, as the size of the network grows, the amount of
information generated by each transmitter decays at the same speed as the
per-node throughput does.  Hence, contrary to the conclusions suggested
in~\cite{GuptaK:00}, designers of these networks should be {\em encouraged}
to consider very large numbers of nodes, for doing so may result in
improved quality of the signals reconstructed at the receivers, and it
may also make more economic sense.

\bigskip
\noindent {\bf Acknowledgements.}  The author would like to thank
Toby Berger, for much needed encouragement and guidance provided
at difficult times; Anna Scaglione, for discussions which resulted
in a solution to a toy problem closely related to this
one~\cite{ScaglioneS:03}; Martin Vetterli, for discussions on the
work of Gupta and Kumar~\cite{GuptaK:00} that greatly contributed
to his understanding of that work; and the anonymous referees, for
their most insightful questions and constructive feedback, which
led to a much improved manuscript.  The author also benefited from
several conversations with V.\ A.\ Vaishampayan and N.\ J.\ A.\ Sloane,
on quantization theory and lattices, in the context of some previous
work~\cite{VaishampayanSS:01}.

\pagebreak
\appendix

\subsection{Bounding $\beta$}
\label{app:trivial1}

Recall from Section~\ref{sec:average-error},
\[
  \beta\;\;\stackrel{\Delta}{=}\;\; \mbox{$\frac 1 n$}
        \sum_{\lambda\in s\kappa(\Lambda)\backslash\{{\bf 0}\}}
        \int_{{\bf x}\in V[\lambda:s\kappa(\Lambda)]}
        ||{\bf x}-\big(\lambda+\gamma_k({\bf x})\big)||^2 f_{X|Y}({\bf x}|\xi)
        \mbox{d}{\bf x},
\]
for any $\xi\in V[{\bf 0}:s\Lambda]$.  Our goal next is to give an
estimate for $\beta$.

Since each term of the sum is positive, we have
a trivial lower bound: $\beta \geq 0$.  As for an upper bound:
\begin{eqnarray}
\beta
  & \stackrel{(a)}{=} & \mbox{$\frac 1 n$}
        \sum_{\lambda\in s\kappa(\Lambda)\!\setminus\{0\}}
        \int_{V[\lambda:s\kappa(\Lambda)]}
        ||{\bf x}-\big(\lambda+\gamma_k({\bf x})\big)||^2
        \frac{1}{[2\pi\sigma_X^2(1-\rho^2)]^{\frac{n}{2}}} \;
        e^{-\frac{n}{2(1-\rho^2)}||\frac{1}{\sigma_X}{\bf x}
                                   -\frac{\rho}{\sigma_Y}{\xi}||^2}
        \mbox{d\bf x}
        \nonumber \\
  & \stackrel{(b)}{\leq} & \mbox{$\frac 1 n$}
        \sum_{\lambda\in s\kappa(\Lambda)\!\setminus\{0\}}
        \int_{V[\lambda:s\kappa(\Lambda)]}
        ||{\bf x}||^2
        \frac{1}{[2\pi\sigma_X^2(1-\rho^2)]^{\frac{n}{2}}} \;
        e^{-\frac{n}{2(1-\rho^2)}||\frac{1}{\sigma_X}{\bf x}
                                   -\frac{\rho}{\sigma_Y}{\xi}||^2}
        \mbox{d\bf x}
        \nonumber \\ & & \mbox{\hspace{2cm}} +
        \int_{V[\lambda:s\kappa(\Lambda)]}
        ||\lambda+\gamma_k({\bf x})||^2
        \frac{1}{[2\pi\sigma_X^2(1-\rho^2)]^{\frac{n}{2}}} \;
        e^{-\frac{n}{2(1-\rho^2)}||\frac{1}{\sigma_X}{\bf x}
                                   -\frac{\rho}{\sigma_Y}{\xi}||^2}
        \mbox{d\bf x}
        \nonumber \\
  & \stackrel{(c)}{\approx} & \mbox{$\frac 1 n$}
        \sum_{\lambda\in s\kappa(\Lambda)\!\setminus\{0\}}
        2||\lambda||^2
        \frac{1}{[2\pi\sigma_X^2(1-\rho^2)]^{\frac{n}{2}}} \;
        e^{-\frac{n}{2\sigma_X^2(1-\rho^2)}||\lambda||^2}
        \left(\int_{V[\lambda:s\kappa(\Lambda)]}\mbox{d\bf x}\right)
        \nonumber \\
  & = & \mbox{$\frac 1 n$}
        \frac{2\nu(s\kappa(\Lambda))}{[2\pi\sigma_X^2(1-\rho^2)]^{\frac{n}{2}}}
        \sum_{\lambda\in s\kappa(\Lambda)\!\setminus\{0\}}
        ||\lambda||^2
        \;e^{-\frac{n}{2\sigma_X^2(1-\rho^2)}||\lambda||^2}
        \nonumber \\
  & = & \mbox{$\frac 1 n$}
        \frac{2\nu(s\kappa(\Lambda))}{[2\pi\sigma_X^2(1-\rho^2)]^{\frac{n}{2}}}
        \sum_{\lambda\in \kappa(\Lambda)\!\setminus\{0\}}
        ||s\lambda||^2
        \;e^{-\frac{n}{2\sigma_X^2(1-\rho^2)}||s\lambda||^2}
        \nonumber \\
  & \stackrel{(d)}{=} & \mbox{$\frac 1 n$}
        \frac{2\nu(s\kappa(\Lambda))s^2}
             {[2\pi\sigma_X^2(1-\rho^2)]^{\frac{n}{2}}}
        \sum_{m=1}^\infty N_m(\kappa(\Lambda))
        \;e^{-\frac{s^2n}{2\sigma_X^2(1-\rho^2)}m}
        \label{eq:b2}
\end{eqnarray}
where:
\begin{itemize}
\item[(a)] is just a substitution for the conditional Gaussian distribution;
\item[(b)] follows from the fact that $||a-b||^2 \leq ||a||^2+||b||^2$;
\item[(c)] is because of two reasons: under the assumption that
  sublattice cells are small, we have $||{\bf x}||^2\approx||\lambda||^2$
  (when ${\bf x}\in V[\lambda:s\kappa(\Lambda)]$); and under the further
  assumption that $R$ is large, $||\gamma_k||^2$ is negligible compared
  to $||\lambda||^2$ (when $\lambda\neq{\bf 0}$), and
  $||\xi||^2\approx{\bf 0}$ (when $\xi\in V[{\bf 0}:s\Lambda]$);
\item[(d)] follows from defining $N_m(\kappa(\Lambda))$ as the number of
  points in $\lambda\in \kappa(\Lambda)$ such that
  $||\lambda||^2=m$.\footnote{Note: wlog, we can take norms to be integers.
  If this is not the case, we can always form a (countable) list of all the
  norms that appear in $\kappa(\Lambda)$, and take $m$ to be an index in
  this list.}
\end{itemize}

To find a useful estimate for this sum, we need to bound
$N_m(\kappa(\Lambda))$.  One simple such bound is:
  \[ N_m(\kappa(\Lambda)) \;\; \leq \;\;
       \frac{\mbox{surface of an $n$-dimensional sphere of radius m}}
            {\mbox{volume of an $(n\!-\!1)$-dimensional sphere of radius
                   $\frac{N}{2}$}}.
  \]
This bound follows from the fact that the highest density of lattice
points on the surface of a sphere cannot be higher than if we assume
a perfect tessellation of this $(n\!-\!1)$-dimensional surface into
$(n\!-\!1)$-dimensional spheres whose radius is $\frac{1}{2}$ of the
smallest separation between sublattice points.  Using standard
formulas~\cite{neil:splag}, we find that
\[ N_m(\kappa(\Lambda))
     \;\; \leq \;\; \frac{c_n m^{n-1}}{d_n \left(\frac{N}{2}\right)^{n-1}}
     \;\; = \;\; e_n m^{n-1},
\]
for appropriate constants $c_n$ and $d_n$, and
$e_n \triangleq \frac{c_n}{d_n(\frac N 2)^{n-1}}$.  Therefore,
\begin{eqnarray}
\beta
  & \stackrel{(a)}{\leq} & \mbox{$\frac 1 n$}
        \frac{2\nu(s\kappa(\Lambda))e_ns^2}
             {[2\pi\sigma_X^2(1-\rho^2)]^{\frac{n}{2}}}
        \sum_{m=1}^\infty m^{n-1}
        \;e^{-\frac{s^2n}{2\sigma_X^2(1-\rho^2)}m}
        \nonumber \\
  & = & \mbox{$\frac 1 n$}
        \frac{2\nu(s\kappa(\Lambda))e_ns^2}
             {[2\pi\sigma_X^2(1-\rho^2)]^{\frac{n}{2}}}
        \sum_{m=1}^\infty
        e^{-\frac{s^2n}{2\sigma_X^2(1-\rho^2)}m+(n-1)\log(m)}
        \nonumber \\
  & \stackrel{(b)}{=} & \mbox{$\frac 1 n$}
        \frac{2\nu(s\kappa(\Lambda))e_ns^2}
             {[2\pi\sigma_X^2(1-\rho^2)]^{\frac{n}{2}}}
        \left(-1+\sum_{m=0}^\infty
                 \left(e^{-\frac{s^2n}{2\sigma_X^2(1-\rho^2)}
                          +\frac{(n-1)\log(m)}{m}}\right)^m\right)
        \nonumber \\
  & \stackrel{(c)}{\leq} & \mbox{$\frac 1 n$}
        \frac{2\nu(s\kappa(\Lambda))e_ns^2}
             {[2\pi\sigma_X^2(1-\rho^2)]^{\frac{n}{2}}}
        \left(-1+\sum_{m=0}^\infty
                 \left(e^{-\frac{s^2}{2\sigma_X^2(1-\rho^2)}}\right)^m\right)
        \nonumber \\
  & \stackrel{(d)}{=} & \mbox{$\frac 1 n$}
        \frac{2\nu(s\kappa(\Lambda))e_ns^2}
             {[2\pi\sigma_X^2(1-\rho^2)]^{\frac{n}{2}}}
        \left(\frac{e^{-\frac{s^2}{2\sigma_X^2(1-\rho^2)}}}
                   {1-e^{-\frac{s^2}{2\sigma_X^2(1-\rho^2)}}}\right)
        \label{eq:b3} \\
  & \stackrel{(e)}{<} & \epsilon
        \nonumber
\end{eqnarray}
where:
\begin{itemize}
\item[(a)] follows from replacing the estimate for $N_m(\kappa(\Lambda))$
  in eqn.~(\ref{eq:b2});
\item[(b)] follows from simple manipulations, and defining
  $\frac{\log 0}{0} = 0$;
\item[(c)] follows from observing that
  $\frac{\log m}{m} < \frac{s^2}{2\sigma_X^2(1-\rho^2)}$, for $\rho^2$ close
  enough to 1;
\item[(d)] follows from evaluation of the sum of a power series;
\item[(e)] where this holds for all values of $\rho$ such
  that $\rho_0 < |\rho| < 1$, for a constant $\rho_0$ that depends on
  $\epsilon$ since, from~(\ref{eq:choice-s}), we have
  $s/\big(\sigma_X\sqrt{1-\rho^2}\big)\to\infty$, thus convergence is
  exponential in $\rho$.
\end{itemize}
Thus, $0\leq \beta < \epsilon$, for all $\epsilon > 0$ and all $|\rho|$
close enough to 1.  Hence, eqn.~(\ref{eq:b3}) defines an asymptotically
good estimate of $\beta$.

\pagebreak

\begin{biography}{Sergio D.\ Servetto}
  was born in Argentina, on January 18, 1968.  He
  received a Licenciatura en Inform\'atica from Universidad Nacional
  de La Plata (UNLP, Argentina) in 1992, and the M.Sc. degree in
  Electrical Engineering and the Ph.D. degree in Computer Science from
  the University of Illinois at Urbana-Champaign (UIUC), in 1996 and
  1999.  Between 1999 and 2001, he worked at the \'Ecole Polytechnique
  F\'ed\'erale de Lausanne (EPFL), Lausanne, Switzerland.  Since Fall
  2001, he has been an Assistant Professor in the School of Electrical
  and Computer Engineering at Cornell University, and a member of the
  fields of Applied Mathematics and Computer Science.  He was the
  recipient of the 1998 Ray Ozzie Fellowship, given to ``outstanding
  graduate students in Computer Science,'' and of the 1999 David J.
  Kuck Outstanding Thesis Award, for the best doctoral dissertation
  of the year, both from the Dept.\ of Computer Science at UIUC.  He
  was also the recipient of a 2003 NSF CAREER Award.  His research
  interests are centered around information theoretic aspects of
  networked systems, with a current emphasis on problems that arise
  in the context of large-scale sensor networks.
\end{biography}

\end{document}